\documentclass[11pt]{article}
\usepackage[top=1in,bottom=1in,left=1in,right=1in]{geometry}
\usepackage{natbib}

\usepackage[unicode=true,bookmarks=true,bookmarksnumbered=false,bookmarksopen=false,breaklinks=false,pdfborder={0 0 1},backref=false,colorlinks=true]{hyperref}
\hypersetup{citecolor=blue,urlcolor=blue}
\usepackage{url,color,xr,xspace,enumerate,paralist}
\usepackage{amsmath,amsfonts,amssymb,mathrsfs,amscd,mathtools,bm}

\usepackage{amsmath}
\usepackage{mathrsfs}
\usepackage{amssymb}
\usepackage{amsthm}
\usepackage[normalem]{ulem}

\newtheorem{theorem}{Theorem}
\newtheorem{lemma}{Lemma}

\newtheorem{definition}{Definition}

\providecommand{\tabularnewline}{\\}
\usepackage{float,rotfloat}
\providecommand{\tabularnewline}{\\}
\floatstyle{ruled}
\usepackage{algpseudocode,algorithm}
\algnewcommand\algorithmicinput{\textbf{Input}:}
\algnewcommand\algorithmicoutput{\textbf{Output}:}
\algnewcommand\INPUT{\item[\algorithmicinput]}
\algnewcommand\OUTPUT{\item[\algorithmicoutput]}
\usepackage{graphicx}
\usepackage[caption=false]{subfig} 

\usepackage[normalem]{ulem}
\usepackage{multirow}

\usepackage{tabularx}
\newcolumntype{L}[1]{>{\raggedright\arraybackslash}p{#1}}
\newcolumntype{C}[1]{>{\centering\arraybackslash}p{#1}}
\newcolumntype{R}[1]{>{\raggedleft\arraybackslash}p{#1}}

\newcommand{\bX}{\mathbf{X}}

\newcommand{\bY}{\mathbf{Y}}

\newcommand{\bW}{\mathbf{W}}

\newcommand{\bM}{\mathbf{M}}

\def\sgn{\operatorname{sgn}}

\newcommand*{\affaddr}[1]{#1} 
\newcommand*{\affmark}[1][*]{\textsuperscript{#1}}

\title{Multimodal Neuroimaging Data Integration and Pathway Analysis}

\author{%
    Yi Zhao\affmark[1], Lexin Li\affmark[2], and Brian S. Caffo\affmark[1] \\
    \affaddr{\affmark[1]Department of Biostatistics, Johns Hopkins Bloomberg School of Public Health} \\
    \affaddr{\affmark[2]Department of Biostatistics and Epidemiology \& Helen Wills Neuroscience Institute, \\
    University of California at Berkeley} \\
}

\date{}

\begin{document}





\maketitle

\thispagestyle{empty}

\begin{abstract}
With fast advancements in technologies, the collection of multiple types of measurements on a common set of subjects is becoming routine in science. Some notable examples include multimodal neuroimaging studies for the simultaneous investigation of brain structure and function, and multi-omics studies for combining genetic and genomic information. Integrative analysis of multimodal data allows scientists to interrogate new mechanistic questions. However, the data collection and generation of integrative hypotheses is outpacing available methodology for joint analysis of multimodal measurements. In this article, we study high-dimensional multimodal data integration in the context of mediation analysis. We aim to understand the roles different data modalities play as possible mediators in the pathway between an exposure variable and an outcome. We propose a mediation model framework with two data types serving as separate sets of mediators, and develop a penalized optimization approach for parameter estimation. We study both the theoretical properties of the estimator through an asymptotic analysis, and its finite-sample performance through simulations. We illustrate our method with a multimodal brain pathway analysis having both structural and functional connectivities as mediators in the association between sex and language processing. 
\bigskip
\end{abstract}



\clearpage

\section{Introduction}
\label{sec:introduction}

Neuroimaging technology is ever expanding with new imaging measurements, such as anatomical magnetic resonance imaging (MRI), functional magnetic resonance imaging (fMRI), diffusion tensor imaging (DTI), positron emission tomography (PET), among many others. These imaging techniques provide crucial tools to facilitate our understanding of brain structure and function, as well as their associations with numerous neurological disorders. Most modern MRI studies are multimodal, in the sense that several types of MRI measurements are collected on the subjects in a scanning session, as different measurements can be obtained from nuclear magnetic resonance in a single scanner. Other more ambitious studies collect multiple imaging measurements for the same group of subjects from different scanners, such as MRI and PET. Intuitively, integration of such diverse, but scientifically complementary neuroimaging information would strengthen our understanding of brain. Although there have been some studies on integrative analysis of multimodal neuroimaging data \citep{zhang2011multimodal, dai2012discriminative, uludaug2014general, lili2018integrative}, many questions remain open, and call for the development of new statistical methodology. 

In this article, we study multimodal data integration in the context of mediation analysis. Mediation analysis seeks to identify and explain the mechanism, or path, that underlies an observed relationship between a treatment or exposure variable and an outcome variable, through the inclusion of an intermediary variable, known as a mediator. Such an analysis is a generalization of path analysis, and represents the starting point for many mechanistic studies. It was originally developed in the psychometric and behavioral sciences literature \citep{baron1986moderator}, but has been extensively studied in the statistics literature \citep[see, e.g.,][among many others]{pearl2001direct, van2008direct, wang2013estimation, vanderweele2014mediation, huang2016hypothesis}. Recently, mediation analysis has received increased attention in neuroimaging analysis to understand the roles of brain structure and function as possible mediators between an exposure variable and some cognitive or behavioral outcome \citep{caffo2007brain, wager2009brain, atlas2010brain, lindquist2012functional, zhao2016pathway, chen2017high, zhao2018sparsepc}. However, all existing work focused on a single imaging modality as the mediator. Herein, we consider the more complex problem of multiple high-dimensional imaging modalities as mediators. 

Our motivation lies in a study of how brain structure and function mediate the relationship between sex and language processing behavior. Sex differences in language processing behavior have been consistently observed \citep[see][for a review]{pinker2007stuff}. Numerous studies have noted sex differences in structure and function of brain regions related to language \citep[e.g.,][]{shaywitz1995sex, kansaku2000sex}. To further investigate this problem, we study a dataset from the Human Connectome Project. It includes a set of $n=136$ young adult participants, of whom 65 are females and 71 males. Each participant went through a battery of cognitive and behavioral tests. We consider the picture vocabulary test, a measure of language processing behavior, as our outcome variable. Each participant also took imaging scanning, including both a DTI and a resting-state fMRI scan. DTI is an MRI technique that measures the diffusion of water molecules, an indirect measure of white matter connectivity, as water diffuses anisotropically along whiter matter fiber bundles. Meanwhile, fMRI is an MRI technique that measures blood oxygen level, which in turn serves as a surrogate measure of brain neural activity. Whereas DTI measures brain structural connectivity, fMRI measures functional connectivity. After preprocessing, each DTI scan is summarized in the form of a 531-dimensional vector, and each fMRI scan as a 917-dimensional vector. We explain in more detail about the DTI and fMRI imaging preprocessing in Section \ref{sec:real}. Logically, brain structural and functional connectivity must be associated. Hebb's law \citep{hebb2005organization} formalizes this notion, observing that distinct brain areas that have communicated frequently are more likely to have more direct structural connections. More recent research also suggests that brain structural connectivity regulates the dynamics of cortical circuits and systems captured by the functional connectivity \citep{sporns2007brain}, and there is increased interest in integrative analysis of structural and functional connectivities \citep{higgins2018integrative}. The goal of our study is to integrate brain structural and functional connectivity to identify brain pathways that associate sex with language behavior. We also comment that, although motivated by a multimodal neuroimaging problem, our method is equally applicable to a large variety of multimodal data, e.g., multi-omics data \citep{shen2013sparse,richardson2016statistical}.

\begin{figure}
\begin{center}
\begin{tabular}{c}
\vspace{-0.5in}
\includegraphics[width=5.0in,height=3.8in]{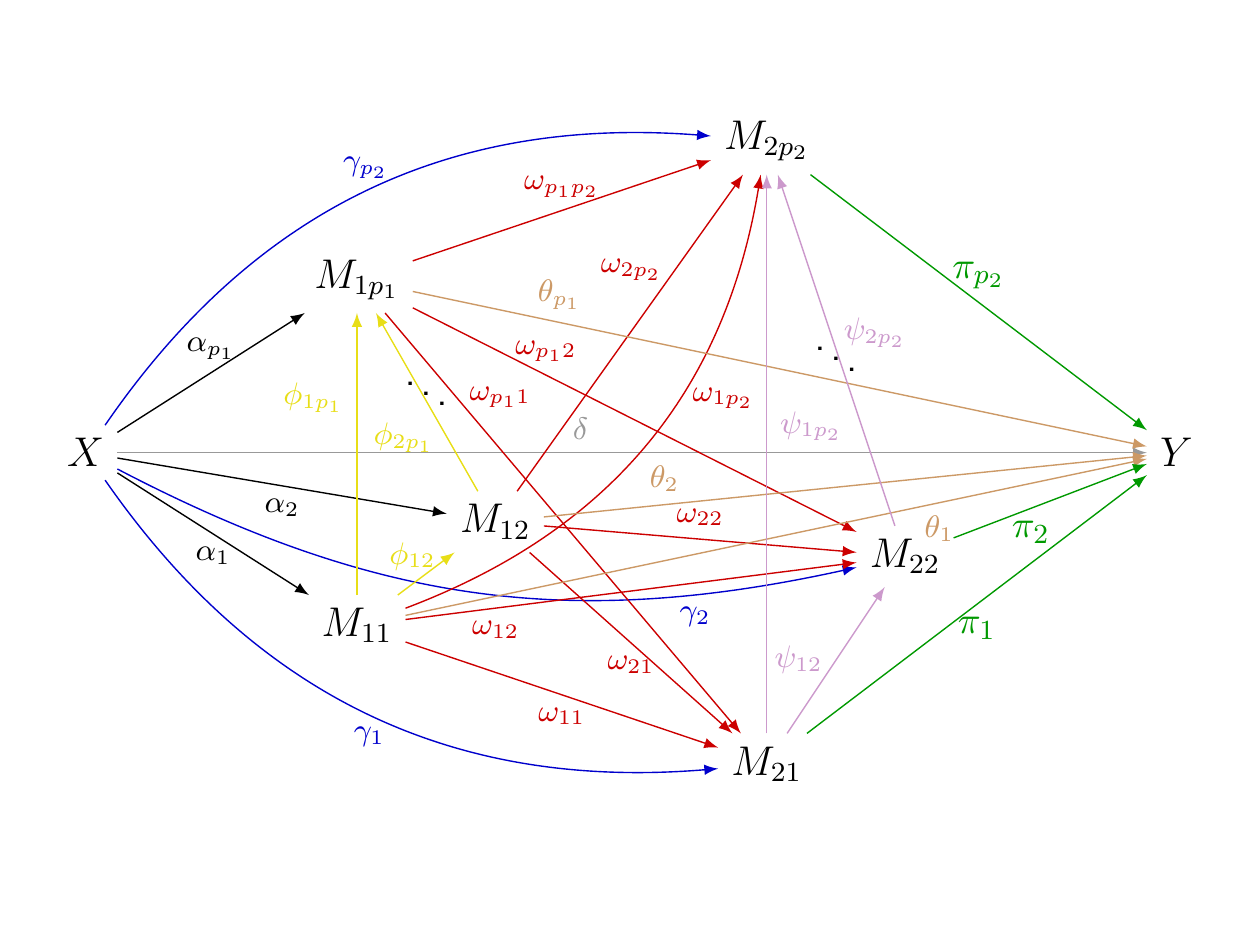} \\ 
Sequential model \\
\vspace{-0.5in}
\includegraphics[width=5.0in,height=3.8in]{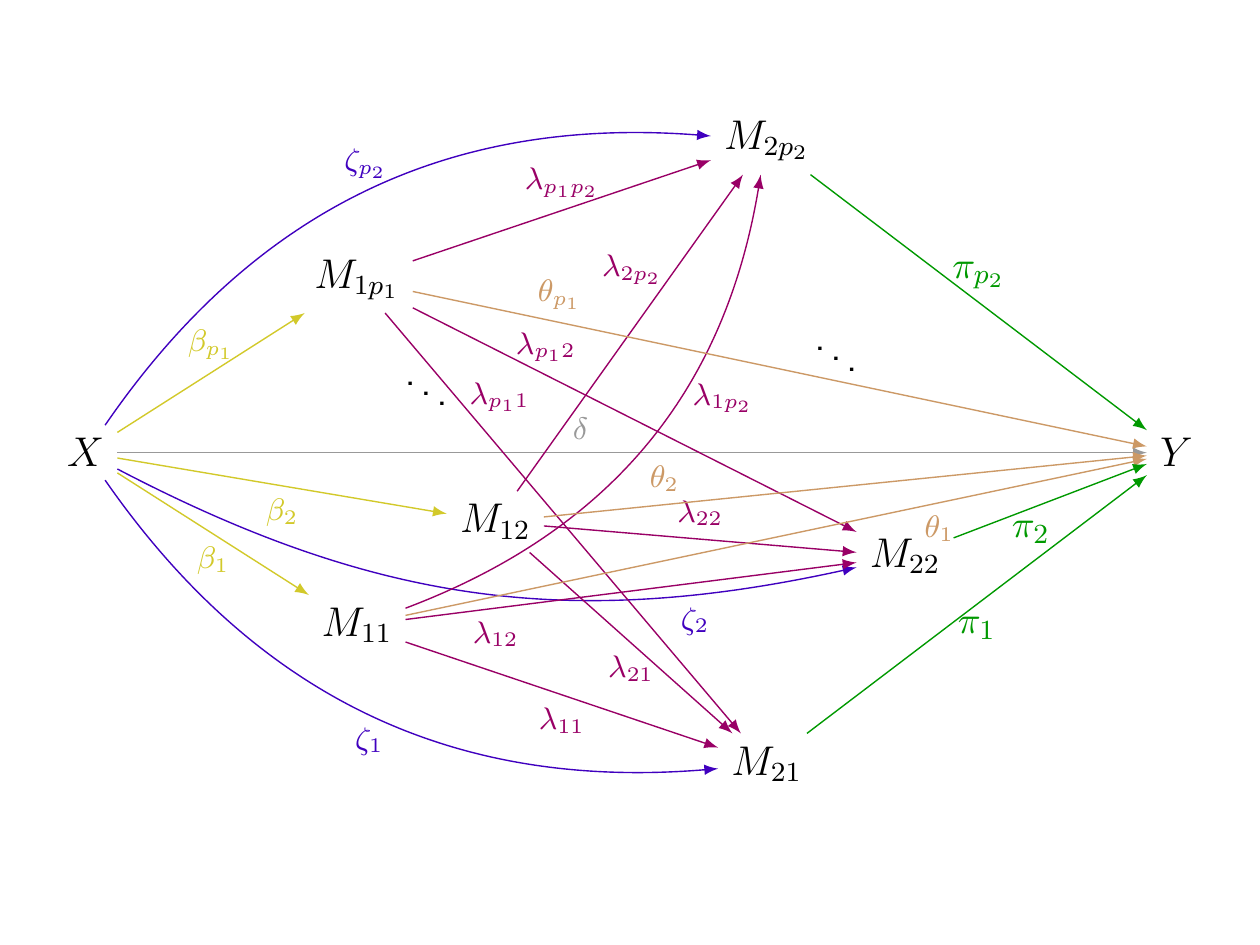} \\
Marginal model
\end{tabular}
\end{center}
\caption{\label{fig:diagram}The diagram of (a) the sequential model, and (b) the proposed marginal model with two sets of mediators. $X$ is the treatment/exposure variable, $\bM_1 = (M_{11},\dots,M_{1p_{1}})^\top$ consists of the first set of mediators, $\bM_2 = (M_{21},\dots,M_{2p_{2}})^\top$ consists of the second set of mediators, and $Y$ is the outcome variable.}
\end{figure}

Toward our study goal, we consider a mediation model framework as depicted in Figure \ref{fig:diagram}. Let $X$ denote the exposure or treatment variable, and $Y$ the outcome variable. In our case, $X$ is the participant's sex, and $Y$ is the picture vocabulary test score. Let $\bM_1 = (M_{11},\dots,M_{1p_{1}})^\top$ denote the set of potential mediators from the first modality, and $\bM_2 = (M_{21},\dots,M_{2p_{2}})^\top$ the set of potential mediators from the second modality. In our case, the structural connectivity measures from DTI are $\bM_1$, and the functional connectivity measures from fMRI are $\bM_2$, with $p_1 = 531$ and $p_2 = 917$. This order of the two sets of mediators is determined by the prior knowledge that structural connectivity shapes and constrains functional connectivity \citep{hagmann2008mapping, honey2009predicting}. We seek to uncover various pathway effects between $X$ and $Y$, including: (i) the indirect effect through some elements of $\bM_1$ but not the rest of $\bM_1$ nor $\bM_2$, (ii) the indirect effect through some elements of $\bM_2$ but not the rest of $\bM_2$ nor $\bM_1$, (iii) the indirect effect through some elements of both $\bM_1$ and $\bM_2$, and (iv) the direct effect of $X$ on $Y$ through neither $\bM_1$ nor $\bM_2$. 

We begin with a \emph{sequential model} as depicted in Figure~\ref{fig:diagram} (a), which captures the pathway effects of interest. However, it faces two immediate challenges. First, in our study, the numbers of the mediators, $p_1 = 531$ and $p_2 = 917$, both far exceed the sample size, $n=136$. To address this issue, we introduce a Lasso-type penalty \citep{tibshirani1996regression} to induce a sparse estimate of the pathway effects. Second, the ordering of the mediators within each modality is unknown. As a result, there are an intractable number of possible combinations of mediation pathways within each modality. To address this issue, we adopt the idea of \citet{zhao2016pathway} and consider a \emph{marginal model}, as depicted in Figure~\ref{fig:diagram} (b). That is, we do not aim to delineate the underlying relationships among the mediators \emph{within} the same modality. Instead, we aim to reveal the roles of the mediators \emph{between} the two sets of modalities in the treatment-outcome pathway. We achieve this by introducing the notion of pathway effect of the mediators, which we formally define in Section \ref{sec:model}, and we argue this is probably of a greater practical interest in scientific applications. We employ correlated errors to account for the dependence among the mediators within each modality, and relax the ordering of the mediators within the same modality. 
Our model, although motivated by \citet{zhao2016pathway}, is distinct in several ways. First, while \citet{zhao2016pathway} tackled the single modality pathway analysis, we focus on the multimodality scenario, for which there is no existing solution. Given the strong demand for this type of multimodal mediation analysis in brain imaging as well as many other applications, we offer a timely solution to this important family of scientific problems. Second, even though a straightforward extension conceptually, our multimodal pathway analysis is technically much more involved than the single modality analysis of \citet{zhao2016pathway}, as we describe in detail later. 

The rest of the paper is organized as follows. Section \ref{sec:model} describes our model, and Section \ref{sec:method} develops the penalized optimization function and the associated estimation algorithm. Section \ref{sec:asmp} studies the asymptotic properties of our estimator. Section \ref{sec:sim} presents the simulation study, and Section \ref{sec:real} revisits our motivating multimodal brain imaging example.  Section \ref{sec:discuss} gives a discussion. The supporting information collects all technical proofs and some additional results.

\section{Model}
\label{sec:model}

In this section, we describe and compare in detail the sequential model and the marginal model, then formally define various path effects of interest. 

We begin with the sequential model.  For this model, we assume the orderings of effects of all variables in $\bM_1$ and $\bM_2$ are completely known, and without loss of generality, the variables in $\bM_1$ and $\bM_2$ are ordered accordingly. We later relax this assumption in the marginal model. For $n$ independent and identically distributed observations, let $\bX \in \mathbb{R}^{n}$ denote the exposure vector, $\bY \in \mathbb{R}^{n}$ the response vector, $\bM_{1j} \in \mathbb{R}^{n}$, and $\bM_{2k} \in \mathbb{R}^{n}$ the two sets of mediators, $j=1,\ldots,p_1$ and $k=1,\ldots,p_2$. We adopt the linear structural equation modeling (LSEM) framework of \citet{pearl2003causality}, and have that, 
\begin{eqnarray}\label{eq:LSEM_full}
&& \bM_{11} = \bX \alpha_{1}+\boldsymbol{\epsilon}_{1}, \; \cdots, \; \bM_{1p_{1}} = \bX \alpha_{p_{1}}+\sum_{j=1}^{p_{1}-1}\bM_{1j}\phi_{jp_{1}}+\boldsymbol{\epsilon}_{p_{1}}, \nonumber \\
&&\bM_{21} = \bX \gamma_{1}+\sum_{j=1}^{p_{1}}\bM_{1j}\omega_{j1}+\boldsymbol{\eta}_{1}, \; \cdots, \; \bM_{2p_{2}} = \bX \gamma_{p_{2}}+\sum_{j=1}^{p_{1}}\bM_{1j}\omega_{jp_{2}}+\sum_{k=1}^{p_{2}-1}\bM_{2k}\psi_{kp_{2}}+\boldsymbol{\eta}_{p_{2}}, \nonumber \\
&&\bY = \bX\delta+\sum_{j=1}^{p_{1}}\bM_{1j}\theta_{j}+\sum_{k=1}^{p_{2}}\bM_{2k}\pi_{k}+\boldsymbol{\xi}, 
\end{eqnarray}
where $\alpha_j$, $\phi_{jj'}$, $\gamma_k$, $\omega_{jk}$, $\psi_{kk'}$, $\delta$, $\theta_j$, $\pi_k$ are all scalar coefficients, and $\boldsymbol{\epsilon}_{j} \in \mathbb{R}^{n}$, $\boldsymbol{\eta}_{k} \in \mathbb{R}^{n}$, $\boldsymbol{\xi} \in \mathbb{R}^{n}$ are independent normal random errors with zero means, $j, j' = 1,\ldots,p_1, k, k' = 1, \ldots, p_2$. Furthermore, the error term $\boldsymbol{\epsilon}_{j}$ is independent of $\bX, \bM_1$, $\boldsymbol{\eta}_{k}$ and $\boldsymbol{\xi}$ are independent of $\bX, \bM_{1}, \bM_2$. We assume the data are centered at zero and thus drop the intercept terms. We depict model \eqref{eq:LSEM_full} in Figure~\ref{fig:diagram}(a). Next we stack the coefficients together, in that 
\begin{eqnarray*}
\boldsymbol{\alpha} = \left( \alpha_{1}, \cdots, \alpha_{p_{1}} \right)_{1 \times p_1}, \;
\boldsymbol{\gamma} = \left(\gamma_{1}, \cdots, \gamma_{p_{2}} \right)_{1 \times p_2}, \;
\boldsymbol{\theta} = \left( \theta_{1}, \cdots, \theta_{p_{1}} \right)^\top_{p_1 \times 1}, \;
\boldsymbol{\pi} = \left( \pi_{1}, \cdots, \pi_{p_{2}} \right)^\top_{p_2 \times 1}, 
\end{eqnarray*}
{\small 
\begin{eqnarray*}
    \boldsymbol{\Phi}=\begin{pmatrix}
        0 & \phi_{12} & \cdots & \phi_{1p_{1}} \\
        & \ddots & \ddots & \vdots \\
        & & \ddots & \phi_{p_{1}-1,p_{1}} \\
        & & & 0
    \end{pmatrix}_{p_1 \times p_1}, \;\;
    \boldsymbol{\Omega}=\begin{pmatrix}
        \omega_{11} & \cdots & \omega_{1p_{2}} \\
        \vdots & \ddots & \vdots \\
        \omega_{p_{1}p_{1}} & \cdots & \omega_{p_{1}p_{2}}
    \end{pmatrix}_{p_1 \times p_2}, \;\; 
    \boldsymbol{\Psi}=\begin{pmatrix}
        0 & \psi_{12} & \cdots & \psi_{1p_{2}} \\
        & \ddots & \ddots & \vdots \\
        & & \ddots & \psi_{p_{2}-1,p_{2}} \\
        & & & 0
    \end{pmatrix}_{p_2 \times p_2}.
\end{eqnarray*}
}
\noindent
We also stack the mediators and the error terms together, in that $\bM_1 = (\bM_{11}, \ldots, \bM_{1p_1}) \in \mathbb{R}^{n \times p_1}$, $\bM_2 = (\bM_{21}, \ldots, \bM_{2p_2}) \in \mathbb{R}^{n \times p_2}$, $\boldsymbol{\epsilon} = (\boldsymbol{\epsilon}_1, \ldots, \boldsymbol{\epsilon}_{p_1}) \in \mathbb{R}^{n \times p_1}$, $\boldsymbol{\eta} = (\boldsymbol{\eta}_1, \ldots, \boldsymbol{\eta}_{p_2}) \in \mathbb{R}^{n \times p_2}$.  Then we can rewrite model \eqref{eq:LSEM_full} in a matrix form, 
\begin{eqnarray}\label{eq:LSEM_full_mat}
 & &  \begin{pmatrix}
        \bM_{1} & \bM_{2} & \bY
    \end{pmatrix}  =  \begin{pmatrix}
        \bX & \bM_{1} & \bM_{2}
    \end{pmatrix}\begin{pmatrix}
        \boldsymbol{\alpha} & \boldsymbol{\gamma} & \delta \\
        \boldsymbol{\Phi} & \boldsymbol{\Omega} & \boldsymbol{\theta} \\
        \boldsymbol{0} & \boldsymbol{\Psi} & \boldsymbol{\pi}
    \end{pmatrix}+\begin{pmatrix}
        \boldsymbol{\epsilon} & \boldsymbol{\eta} & \boldsymbol{\xi}
    \end{pmatrix},
\end{eqnarray}
where $\mathrm{vec}(\boldsymbol{\epsilon}) \sim \mathcal{N}(\boldsymbol{\mathrm{0}},\boldsymbol{\Xi}_{1}\otimes\boldsymbol{\mathrm{I}}_n)$, $\mathrm{vec}(\boldsymbol{\eta}) \sim \mathcal{N}(\boldsymbol{\mathrm{0}},\boldsymbol{\Xi}_{2}\otimes\boldsymbol{\mathrm{I}}_n)$, $\boldsymbol{\xi}\sim\mathcal{N}(\boldsymbol{\mathrm{0}},\sigma^{2}\boldsymbol{\mathrm{I}}_n)$, $\boldsymbol{\Xi}_{1} \in \mathbb{R}^{p_{1} \times p_{1}}$, $\boldsymbol{\Xi}_{2} \in \mathbb{R}^{p_{2} \times p_{2}}$ are the covariance matrices, and $\boldsymbol{\mathrm{I}}_n$ is the identity matrix with dimension $n$. Model~\eqref{eq:LSEM_full} fully characterizes the dependencies among all the variables, including the treatment, the mediators, and the outcome. The model errors $\boldsymbol{\epsilon}$, $\boldsymbol{\eta}$ and $\boldsymbol{\xi}$ are assumed to be mutually independent, and $\boldsymbol{\Xi}_{1}$ and $\boldsymbol{\Xi}_{2}$ are diagonal matrices. 

Model \eqref{eq:LSEM_full} requires the knowledge of the ordering of the mediators within each modality, which is generally unknown in practice. To circumvent this challenge, we consider an alternative model of the form, 
\begin{eqnarray}\label{eq:LSEM_reduced}
\bM_{1j} &=& \bX \beta_j + \boldsymbol{\varepsilon}_j, \;\; j=1, \ldots, p_1, \nonumber \\
\bM_{2k} &=& \bX \zeta_k + \sum_{j=1}^{p_{1}}\bM_{1j} \lambda_{jk} + \boldsymbol{\vartheta}_k, \;\; k=1, \ldots, p_2, \nonumber \\
\bY & = & \bX\delta+\sum_{j=1}^{p_{1}}\bM_{1j}\theta_{j}+\sum_{k=1}^{p_{2}}\bM_{2k}\pi_{k}+\boldsymbol{\xi}, 
\end{eqnarray}
where $\beta_j$, $\zeta_k$, $\lambda_{jk}$ are scalar coefficients, $\delta$, $\theta_j$, $\pi_k$, $\boldsymbol{\xi}$ are the same as defined in model \eqref{eq:LSEM_full}, and $\boldsymbol{\varepsilon}_{j} \in \mathbb{R}^{n}$, $\boldsymbol{\vartheta}_{k} \in \mathbb{R}^{n}$ are normal random errors with zero means, $j=1,\dots,p_{1}$, $k=1,\dots,p_{2}$. The error term $\boldsymbol{\varepsilon}_{j}$ is independent of $\bX$, and $\boldsymbol{\vartheta}_{k}$ is independent of $\bX, \bM_{1}$. We depict model \eqref{eq:LSEM_reduced} in Figure \ref{fig:diagram} (b). It extends that of \citet{zhao2016pathway} from a single modality of mediators to multiple modalities of mediators. Next we stack the coefficients, in that  
\begin{eqnarray*}
\boldsymbol{\beta} = \left( \beta_{1}, \cdots, \beta_{p_{1}} \right)_{1 \times p_1}, \;
\boldsymbol{\zeta} = \left(\zeta_{1}, \cdots, \zeta_{p_{2}} \right)_{1 \times p_2}, \; 
\boldsymbol{\Lambda}=
\begin{pmatrix}
        \lambda_{11} & \cdots & \lambda_{1p_{2}} \\
        \vdots & \ddots & \vdots \\
        \lambda_{p_{1}p_{1}} & \cdots & \lambda_{p_{1}p_{2}}
\end{pmatrix}_{p_1 \times p_2},
\end{eqnarray*}
and the error terms $\boldsymbol{\varepsilon} = (\boldsymbol{\varepsilon}_1, \ldots, \boldsymbol{\varepsilon}_{p_1}) \in \mathbb{R}^{n \times p_1}$, and $\boldsymbol{\vartheta} = (\boldsymbol{\vartheta}_1, \ldots, \boldsymbol{\vartheta}_{p_2}) \in \mathbb{R}^{n \times p_2}$. Then we can again rewrite model \eqref{eq:LSEM_reduced} in a matrix form, 
\begin{eqnarray}\label{eq:LSEM_full_mat}
 & &  \begin{pmatrix}
        \bM_{1} & \bM_{2} & \bY
    \end{pmatrix}  =  \begin{pmatrix}
        \bX & \bM_{1} & \bM_{2}
    \end{pmatrix}\begin{pmatrix}
        \boldsymbol{\beta} & \boldsymbol{\zeta} & \delta \\
        \boldsymbol{0} & \boldsymbol{\Lambda} & \boldsymbol{\theta} \\
        \boldsymbol{0} & \boldsymbol{0} & \boldsymbol{\pi}
    \end{pmatrix}+\begin{pmatrix}
        \boldsymbol{\varepsilon} & \boldsymbol{\vartheta} & \boldsymbol{\xi}
    \end{pmatrix}, 
 \end{eqnarray}
where $\mathrm{vec}(\boldsymbol{\varepsilon}) \sim \mathcal{N}(\boldsymbol{\mathrm{0}},\boldsymbol{\Sigma}_{1}\otimes\boldsymbol{\mathrm{I}}_n)$, $\mathrm{vec}(\boldsymbol{\vartheta}) \sim \mathcal{N}(\boldsymbol{\mathrm{0}},\boldsymbol{\Sigma}_{2}\otimes\boldsymbol{\mathrm{I}}_n)$, and $\boldsymbol{\xi}\sim\mathcal{N}(\boldsymbol{\mathrm{0}},\sigma^{2}\boldsymbol{\mathrm{I}}_n)$.

Comparing the marginal model \eqref{eq:LSEM_reduced} to the sequential model \eqref{eq:LSEM_full}, for $\bM_{1}$, model \eqref{eq:LSEM_reduced} can be viewed as a total effect model of $\bX$~\citep{imai2010identification}, which avoids the explicit modeling of the relationship among the variables in $\bM_{1}$. A similar relation holds for $\bM_{2}$. Consequently, model \eqref{eq:LSEM_reduced} does not require the knowledge of the within-modality mediator ordering. Moreover, the model parameters between the two models satisfy: 
\begin{eqnarray*}
\boldsymbol{\beta}=\boldsymbol{\alpha}(\boldsymbol{\mathrm{I}}-\boldsymbol{\Phi})^{-1}, \quad \boldsymbol{\zeta}=\boldsymbol{\gamma}(\boldsymbol{\mathrm{I}}-\boldsymbol{\Psi})^{-1}, \quad \boldsymbol{\Lambda}=\boldsymbol{\Omega}(\boldsymbol{\mathrm{I}}-\boldsymbol{\Psi})^{-1}, \\
\boldsymbol{\Sigma}_{1}=(\boldsymbol{\mathrm{I}}-\boldsymbol{\Phi}^\top)^{-1}\boldsymbol{\Xi}_{1}(\boldsymbol{\mathrm{I}}-\boldsymbol{\Phi})^{-1}, \quad \boldsymbol{\Sigma}_{2}=(\boldsymbol{\mathrm{I}}-\boldsymbol{\Psi}^\top)^{-1}\boldsymbol{\Xi}_{2}(\boldsymbol{\mathrm{I}}-\boldsymbol{\Psi})^{-1},
\end{eqnarray*}
where $\boldsymbol{\Phi}$ and $\boldsymbol{\Psi}$ are the weighted adjacency matrices of $\bM_{1}$ and $\bM_{2}$, respectively, which are upper-triangular matrices with zero diagonal elements demonstrating the direct impact between the mediators, and $(\boldsymbol{\mathrm{I}}-\boldsymbol{\Phi})^{-1}$ and $(\boldsymbol{\mathrm{I}}-\boldsymbol{\Psi})^{-1}$ are the influence matrices that reveal the overall influence of one mediator on the other. The term $\boldsymbol{\beta}$, which is the product of the influence matrix and the corresponding treatment effect $\boldsymbol{\alpha}$, summarizes the overall treatment effect on the corresponding mediator $\bM_1$. A similar interpretation applies to $\boldsymbol{\zeta}$. The $(j,k)$th element of $\boldsymbol{\Omega}$ captures the direct impact of $M_{1j}$ on $M_{2k}$, and the corresponding element in $\boldsymbol{\Lambda}$ reveals the overall effect regardless of the underlying relationships among the mediators in $\bM_{2}$. Furthermore, under model \eqref{eq:LSEM_full}, the error terms in $\boldsymbol{\epsilon}$ and $\boldsymbol{\eta}$ are mutually independent. By contrast, under model \eqref{eq:LSEM_reduced}, the error terms in $\boldsymbol{\varepsilon}$  and $\boldsymbol{\vartheta}$ are dependent, due to the influences among the mediators. Therefore, even though model \eqref{eq:LSEM_reduced} does not explicitly model the relationship among the mediators within the same modality, it encapsulates the dependencies among the mediators through the correlations between the error terms. Next we formally define the various pathway effects of interest under model \eqref{eq:LSEM_reduced}. 

\begin{definition}\label{def:patheffect} Under model~\eqref{eq:LSEM_reduced}, considering two treatment/exposure conditions $X = x$ and $X = x^{*}$, we define the following pathway effects of $X$ on the outcome $Y$. 
\begin{enumerate}[(i)]
\item The indirect pathway effect of $X$ through path $X\rightarrow M_{1j}\rightarrow Y$ is: $\mathrm{IE}_{j}^{1}(x,x^{*})=\beta_{j}\theta_{j}(x-x^{*})$, $j=1,\dots,p_{1}$. The total indirect pathway effect of $X$ through $\bM_{1}$ but not through $\bM_{2}$ is: $\mathrm{IE}^{1}(x,x^{*})=\sum_{j=1}^{p_{1}}\beta_{j}\theta_{j}(x-x^{*})$. 
   
\item The indirect pathway effect of $X$ through path $X\rightarrow M_{2k}\rightarrow Y$ is: $\mathrm{IE}_{k}^{2}(x,x^{*})=\zeta_{k}\pi_{k}(x-x^{*})$, $k=1,\dots,p_{2}$. The total indirect pathway effect of $X$ through $\bM_{2}$ but not through $\bM_{1}$ is $\mathrm{IE}^{2}(x,x^{*})=\sum_{k=1}^{p_{2}}\zeta_{k}\pi_{k}(x-x^{*})$. 

\item The indirect pathway effect of $X$ through path $X\rightarrow M_{1j}\rightarrow M_{2k}\rightarrow Y$ is: $\mathrm{IE}_{jk}^{1,2}(x,x^{*})=\beta_{j}\lambda_{jk}\pi_{k}(x-x^{*})$, $j=1,\dots,p_{1}, k=1,\dots,p_{2}$. The total indirect effect of $X$ through both $\bM_{1}$ and $\bM_{2}$ is: $\mathrm{IE}^{1,2}=\sum_{j=1}^{p_{1}}\sum_{k=1}^{p_{2}}\beta_{j}\lambda_{jk}\pi_{k}(x-x^{*})$. 

\item The direct effect of $X$ is: $\mathrm{DE}(x,x^{*})=\delta(x-x^{*})$. 
\end{enumerate}
\end{definition}

\noindent
By Definition~\ref{def:patheffect}, the total effect (TE) of $X$ on $Y$ is decomposed as the sum of the direct effect (DE) and the total indirect effect (IE): 
\begin{eqnarray} \label{eqn:total-effect-decomp}
\mathrm{TE}(x,x^{*}) = \mathrm{DE}(x,x^{*}) + \mathrm{IE}(x,x^{*}) \equiv \mathrm{DE}(x,x^{*})+\mathrm{IE}^{1}(x,x^{*})+\mathrm{IE}^{2}(x,x^{*})+\mathrm{IE}^{1,2}(x,x^{*}).
\end{eqnarray}

\section{Estimation}
\label{sec:method}

Our goal is to estimate the pathway effects defined in Definition~\ref{def:patheffect} under model~\eqref{eq:LSEM_reduced}. To achieve this, we define the objective function, 
\begin{eqnarray*}
\ell(\boldsymbol{\beta},\boldsymbol{\theta},\boldsymbol{\zeta},\boldsymbol{\pi},\boldsymbol{\Lambda},\delta)  & = & \textrm{trace}\left\{ (\bM_{1}-\bX\boldsymbol{\beta})^\top(\bM_{1}-\bX\boldsymbol{\beta}) \right\} \\
& & + \; \textrm{trace}\left\{ (\bM_{2}-\bX\boldsymbol{\zeta}-\bM_{1}\boldsymbol{\Lambda})^\top(\bM_{2}-\bX\boldsymbol{\zeta}-\bM_{1}\boldsymbol{\Lambda}) \right\} \\
& & + \; \left(\bY-\bX\delta-\bM_{1}\boldsymbol{\theta}-\bM_{2}\boldsymbol{\pi}\right)^\top\left(\bY-\bX\delta-\bM_{1}\boldsymbol{\theta}-\bM_{2}\boldsymbol{\pi}\right).
\end{eqnarray*}
This objective function conceptually sets $\boldsymbol{\Sigma}_{1}$ and $\boldsymbol{\Sigma}_{2}$ to be identity matrices. However, this simplification would not affect the consistency of our final estimators as long as all the variables are standardized to unit scale \citep{huber1967behavior, white1980heteroskedasticity}. 

Next we introduce a series of penalty functions and consider the following regularized optimization problem,
\begin{eqnarray*}
&& \textrm{minimize}_{\boldsymbol{\beta},\boldsymbol{\theta},\boldsymbol{\zeta},\boldsymbol{\pi},\boldsymbol{\Lambda},\delta} \; 
\ell(\boldsymbol{\beta},\boldsymbol{\theta},\boldsymbol{\zeta},\boldsymbol{\pi},\boldsymbol{\Lambda},\delta) \\ [5pt]
& & \quad\quad\quad \textrm{ subject to } \sum_{j=1}^{p_{1}}|\beta_{j}\theta_{j}|\leq t_{1}, \sum_{k=1}^{p_{2}}|\zeta_{k}\pi_{k}|\leq t_{2},  \sum_{j=1}^{p_{1}}\sum_{k=1}^{p_{2}}|\beta_{j}\lambda_{jk}\pi_{k}|\leq t_{3}, |\delta|\leq t_{4}, 
\end{eqnarray*}
where $t_{1},t_{2},t_{3},t_{4}\geq 0$ are the regularization parameters. Our penalty functions are not convex. The next lemma presents a convex relaxation through elastic net type penalties \citep{zou2005regularization}. We also observe that, in our study, the mediation effect through both $\bM_{1}$ and $\bM_{2}$ is defined as a three-way product, $\beta_{j}\lambda_{jk}\pi_{k}$. Since $\beta_{j}$ and $\pi_{k}$ are already regularized in the pathway effect through $\bM_{1}$ but not through $\bM_{2}$, i.e., $\beta_j \theta_j$, and that through $\bM_{2}$ but not through $\bM_{1}$, i.e., $\zeta_k \pi_k$, respectively, it suffices to regularize $\lambda_{jk}$ alone.

\begin{lemma}\label{lemma:convex}
For any $\nu_{1},\nu_{2} \in\mathbb{R} \geq 1/2$,
\[
\sum_{j=1}^{p_{1}}\left\{ |\beta_{j}\theta_{j}|+\nu_{1}(\beta_{j}^{2}+\theta_{j}^{2}) \right\}, \quad 
\sum_{k=1}^{p_{2}}\left\{ |\zeta_{k}\pi_{k}|+\nu_{2}(\zeta_{k}^{2}+\pi_{k}^{2}) \right\}, \quad \text{and } 
\sum_{j=1}^{p_{1}}\sum_{k=1}^{p_{2}}|\lambda_{jk}|
\]
are convex functions of $\{\boldsymbol{\beta},\boldsymbol{\theta}\}$, $\{\boldsymbol{\zeta},\boldsymbol{\pi}\}$ and $\boldsymbol{\Lambda}$, respectively. For any $t_{1},t_{2},t_{3}\in\mathbb{R}\geq 0$, there exist $r_{1},r_{2},r_{3}\in\mathbb{R}\geq 0$, such that
\[
\begin{cases}
\sum_{j=1}^{p_{1}}\left\{ |\beta_{j}\theta_{j}|+\nu_{1}(\beta_{j}^{2}+\theta_{j}^{2}) \right\} \leq r_{1} \\
\sum_{k=1}^{p_{2}}\left\{ |\zeta_{k}\pi_{k}|+\nu_{2}(\zeta_{k}^{2}+\pi_{k}^{2}) \right\} \leq r_{2} \\
\sum_{j=1}^{p_{1}}\sum_{k=1}^{p_{2}}|\lambda_{jk}|\leq r_{3} \\
\end{cases} \quad  \Rightarrow  \quad
\begin{cases}
\sum_{j=1}^{p_{1}}|\beta_{j}\theta_{j}|\leq t_{1} \\
\sum_{k=1}^{p_{2}}|\zeta_{k}\pi_{k}|\leq t_{2} \\
\sum_{j=1}^{p_{1}}\sum_{k=1}^{p_{2}}|\beta_{j}\lambda_{jk}\pi_{k}|\leq t_{3}
\end{cases}.
\]    
\end{lemma}

Based on this convex relaxation, we turn to the following optimization problem, 
\begin{eqnarray}\label{eq:objfunc}
\textrm{minimize}_{\boldsymbol{\beta},\boldsymbol{\theta},\boldsymbol{\zeta},\boldsymbol{\pi},\boldsymbol{\Lambda},\delta} \; 
\left\{
\frac{1}{2} \ell(\boldsymbol{\beta},\boldsymbol{\theta},\boldsymbol{\zeta},\boldsymbol{\pi},\boldsymbol{\Lambda},\delta) 
+ P_{1}(\boldsymbol{\beta},\boldsymbol{\theta},\boldsymbol{\zeta},\boldsymbol{\pi}) 
+ P_{2}(\boldsymbol{\beta},\boldsymbol{\theta},\boldsymbol{\zeta},\boldsymbol{\pi}) 
+ P_{3}(\boldsymbol{\Lambda},\delta)
\right\}, 
\end{eqnarray}
where the three penalty functions are of the form, 
\begin{eqnarray*}
P_{1}(\boldsymbol{\beta},\boldsymbol{\theta},\boldsymbol{\zeta},\boldsymbol{\pi}) &=& \kappa_{1} \left[ \sum_{j=1}^{p_{1}} \left\{ |\beta_{j}\theta_{j}|+\nu_{1}(\beta_{j}^{2}+\theta_{j}^{2}) \right\} \right] + \kappa_{2} \left[ \sum_{k=1}^{p_{2}} \left\{ |\zeta_{k}\pi_{k}|+\nu_{2}(\zeta_{k}^{2}+\pi_{k}^{2}) \right\} \right], \\
P_{2}(\boldsymbol{\beta},\boldsymbol{\theta},\boldsymbol{\zeta},\boldsymbol{\pi}) &=& \mu_{1} \sum_{j=1}^{p_{1}}(|\beta_{j}|+|\theta_{j}|)+\mu_{2} \sum_{k=1}^{p_{2}}(|\zeta_{k}|+|\pi_{k}|), \\
P_{3}(\boldsymbol{\Lambda},\delta) &=& \kappa_{3} \sum_{j=1}^{p_{1}}\sum_{k=1}^{p_{2}}|\lambda_{jk}|+\kappa_{4} |\delta|, 
\end{eqnarray*}
with $\nu_{1},\nu_{2} \geq 1/2$, $\kappa_{1},\kappa_{2},\kappa_{3},\kappa_{4} \geq 0$, and $\mu_{1},\mu_{2}\geq 0$ as the tuning parameters. Here $\nu_{1},\nu_{2}$ control the level of convexity relaxiation, $\kappa_{1},\kappa_{2},\kappa_{3},\kappa_{4}$ control the level of penalty on various pathway effects, and $\mu_{1},\mu_{2}$ control the level of sparsity of individual parameters. We note that the tuning parameters $\kappa_{1},\kappa_{2},\kappa_{3},\mu_{1},\mu_{2}$ can also vary with $j$ and $k$. For simplicity, we keep them the same across $1 \leq j \leq p_1$ and $1 \leq k \leq p_2$. 

The objective function in \eqref{eq:objfunc} consists of a differentiable loss function $\ell/{2}$ and an indifferentiable regularization function $(P_{1} + P_{2} + P_{3})$. We next develop an alternating direction method of multipliers \citep[ADMM,][]{boyd2011distributed} to solve \eqref{eq:objfunc}. The ADMM form of the optimization problem \eqref{eq:objfunc} is,
\begin{eqnarray*}
&& \textrm{minimize}_{\boldsymbol{\beta},\boldsymbol{\theta},\boldsymbol{\zeta},\boldsymbol{\pi},\boldsymbol{\Lambda},\delta,
\tilde{\boldsymbol{\beta}},\tilde{\boldsymbol{\theta}},\tilde{\boldsymbol{\zeta}},\tilde{\boldsymbol{\pi}} } \; 
\frac{1}{2}\ell(\boldsymbol{\beta},\boldsymbol{\theta},\boldsymbol{\zeta},\boldsymbol{\pi},\boldsymbol{\Lambda},\delta) + P_{1}(\tilde{\boldsymbol{\beta}},\tilde{\boldsymbol{\theta}},\tilde{\boldsymbol{\zeta}},\tilde{\boldsymbol{\pi}}) + P_{2}(\tilde{\boldsymbol{\beta}},\tilde{\boldsymbol{\theta}},\tilde{\boldsymbol{\zeta}},\tilde{\boldsymbol{\pi}}) + P_{3}(\boldsymbol{\Lambda},\delta), \\
& & \quad\quad\quad \textrm{ subject to } \boldsymbol{\beta}=\tilde{\boldsymbol{\beta}}, \; \boldsymbol{\theta}=\tilde{\boldsymbol{\theta}}, \; \boldsymbol{\zeta}=\tilde{\boldsymbol{\zeta}}, \; \boldsymbol{\pi}=\tilde{\boldsymbol{\pi}},
\end{eqnarray*}
where $\tilde{\boldsymbol{\beta}} \in \mathbb{R}^{1\times p_{1}}$, $\tilde{\boldsymbol{\theta}} \in \mathbb{R}^{p_{1}}$, $\tilde{\boldsymbol{\zeta}} \in \mathbb{R}^{1\times p_{2}}$, and $\tilde{\boldsymbol{\pi}}\in\mathbb{R}^{p_{2}}$ are the newly introduced parameters. Let $\boldsymbol{\Upsilon}=(\boldsymbol{\beta},\boldsymbol{\theta},\boldsymbol{\zeta},\boldsymbol{\pi})$, $\tilde{\boldsymbol{\Upsilon}}=(\tilde{\boldsymbol{\beta}},\tilde{\boldsymbol{\theta}},\tilde{\boldsymbol{\zeta}},\tilde{\boldsymbol{\pi}})$, the augmented Lagrangian function to enforce the constraints is, 
\begin{equation}\label{eq:ADMM_AL}
\frac{1}{2}\ell(\boldsymbol{\Upsilon},\boldsymbol{\Lambda},\delta) + P_{1}(\tilde{\boldsymbol{\Upsilon}}) + P_{2}(\tilde{\boldsymbol{\Upsilon}}) + P_{3}(\boldsymbol{\Lambda},\delta)+\sum_{r=1}^{4}\left( \langle h_{r}(\boldsymbol{\Upsilon},\tilde{\boldsymbol{\Upsilon}}),\boldsymbol{\tau}_{r}\rangle+\frac{\rho}{2}\| h_{r}(\boldsymbol{\Upsilon},\tilde{\boldsymbol{\Upsilon}})\|_{2}^{2}\right),
\end{equation}
where $h_{1}(\boldsymbol{\Upsilon},\tilde{\boldsymbol{\Upsilon}})=\boldsymbol{\beta}-\tilde{\boldsymbol{\beta}}$, $h_{2}(\boldsymbol{\Upsilon},\tilde{\boldsymbol{\Upsilon}})=\boldsymbol{\theta}-\tilde{\boldsymbol{\theta}}$, $h_{3}(\boldsymbol{\Upsilon},\tilde{\boldsymbol{\Upsilon}})=\boldsymbol{\zeta}-\tilde{\boldsymbol{\zeta}}$, $h_{4}(\boldsymbol{\Upsilon},\tilde{\boldsymbol{\Upsilon}})=\boldsymbol{\pi}-\tilde{\boldsymbol{\pi}}$, $\boldsymbol{\tau}_{1},\boldsymbol{\tau}_{2}\in\mathbb{R}^{p_{1}}$, $\boldsymbol{\tau}_{3},\boldsymbol{\tau}_{4}\in\mathbb{R}^{p_{2}}$, and $\rho>0$ is the augmented Lagrangian parameter. We propose to update the parameters in~\eqref{eq:ADMM_AL} iteratively, and summarize our estimation procedure in Algorithm~\ref{alg:ADMM}.

\begin{algorithm}[t!]
\caption{The optimization algorithm for \eqref{eq:ADMM_AL}.}
\begin{algorithmic}[1]
\INPUT $(\bX,\bM_{1},\bM_{2},\bY)$. 
\OUTPUT $\left( \hat{\boldsymbol{\beta}},\hat{\boldsymbol{\theta}},\hat{\boldsymbol{\zeta}},\hat{\boldsymbol{\pi}},\hat{\boldsymbol{\Lambda}},\hat{\delta} \right)$. 

\State \textbf{initialization}: $\left\{ \boldsymbol{\beta}^{(0)},\boldsymbol{\theta}^{(0)},\boldsymbol{\zeta}^{(0)},\boldsymbol{\pi}^{(0)},\boldsymbol{\Lambda}^{(0)},\delta^{(0)},\tilde{\boldsymbol{\beta}}^{(0)},\tilde{\boldsymbol{\theta}}^{(0)},\tilde{\boldsymbol{\zeta}}^{(0)},\tilde{\boldsymbol{\pi}}^{(0)},\boldsymbol{\tau}_{1}^{(0)},\boldsymbol{\tau}_{2}^{(0)},\boldsymbol{\tau}_{3}^{(0)},\boldsymbol{\tau}_{4}^{(0)} \right\}$.

\Repeat
\State update $\boldsymbol{\beta}^{(s+1)} = (\bX^\top\bX+\rho)^{-1} \left\{ \bX^\top\bM_{1}-\boldsymbol{\tau}_{1}^{(s)\top}+\rho\tilde{\boldsymbol{\beta}}^{(s)} \right\}$.

\State update $\boldsymbol{\theta}^{(s+1)} = (\bM_{1}^\top\bM_{1}+\rho\boldsymbol{\mathrm{I}})^{-1} \left[ \bM_{1}^\top \left\{ \bY-\bX\delta^{(s)}-\bM_{2}\boldsymbol{\pi}^{(s)} \right\} - \boldsymbol{\tau}_{2}^{(s)}+\rho\tilde{\boldsymbol{\theta}}^{(s)} \right]$.
    
\State update $\boldsymbol{\zeta}^{(s+1)} = (\bX^\top\bX+\rho)^{-1} \left[ \bX^\top \left\{ \bM_{2}-\bM_{1}\boldsymbol{\Lambda}^{(s)} \right\} - \boldsymbol{\tau}_{3}^{(s)\top}+\rho\tilde{\boldsymbol{\zeta}}^{(s)} \right]$.
    
\State update $\boldsymbol{\pi}^{(s+1)} = (\bM_{2}^\top\bM_{2} + \rho\boldsymbol{\mathrm{I}})^{-1} \left[ \bM_{2}^\top \left\{ \bY-\bX\delta^{(s)}-\bM_{1}\boldsymbol{\theta}^{(s+1)} \right\} - \boldsymbol{\tau}_{4}^{(s)}+\rho\tilde{\boldsymbol{\pi}}^{(s)} \right]$.
    
\State update $\delta^{(s+1)} = (\bX^\top\bX)^{-1} \mathrm{Soft}\left[ \bX^\top \left\{ \bY-\bM_{1}\boldsymbol{\theta}^{(s+1)}-\bM_{2}\boldsymbol{\pi}^{(s+1)} \right\}, \kappa_{4} \right]$.
  
\For {$k$ = 1 to $p_2$}
    \State update $\boldsymbol{\Lambda}_{k}^{(s+1)}$ by solving \eqref{eq:lambda_k}. 
\EndFor
    
\For {$j$ = 1 to $p_1$}
    \State update $\left\{ \tilde{\beta}_{j}^{(s+1)},\tilde{\theta}_{j}^{(s+1)} \right\}$ by solving \eqref{eq:beta_theta_tilde}. 
\EndFor
    
\For {$k$ = 1 to $p_2$}
    \State update $\left\{ \tilde{\zeta}_{k}^{(s+1)},\tilde{\pi}_{k}^{(s+1)} \right\}$ by solving \eqref{eq:zeta_pi_tilde}. 
\EndFor
      
\State update $\boldsymbol{\tau}_{r}^{(s+1)} = \boldsymbol{\tau}_{r}^{(s)}+\rho h_{r}\left\{ \boldsymbol{\Upsilon}^{(s+1)},\tilde{\boldsymbol{\Upsilon}}^{(s+1)} \right\}$, $r=1,\ldots,4$. 

\Until{the objective function converges.}
\end{algorithmic}
\label{alg:ADMM}
\end{algorithm}

A few remarks are in order. The explicit forms of Steps 3--6 of Algorithm~\ref{alg:ADMM} are derived in Section~\ref{appendix:sec:ADMM} of the supporting information. In Step 7, $\textrm{Soft}(a, b)=\sgn(a)\max\{|a|-b,0\}$ is the soft-thresholding function. Steps 8--10 are to update $\boldsymbol{\Lambda}$, one column at a time. Its $k$th column, $k = 1, \ldots, p_2$, can be obtained by,  
\begin{eqnarray}\label{eq:lambda_k}
\textrm{minimize}_{\boldsymbol{\Lambda}_k\in\mathbb{R}^{p_{1}}}~\frac{1}{2}\|\bM_{2k}-\bX\zeta_{k}^{(s+1)}-\bM_{1}\boldsymbol{\Lambda}_k\|_{2}^{2}+\kappa_{3}\|\boldsymbol{\Lambda}_{k}\|_{1}.
\end{eqnarray}
This is a standard Lasso problem with $\left\{ \bM_{2k}-\bX\zeta_{k}^{(s+1)} \right\}$ as the ``outcome" and $\bM_{1}$ as the ``predictor". Steps 11--13 are to update $(\tilde{\boldsymbol{\beta}},\tilde{\boldsymbol{\theta}})$, also one column pair at a time, and the $j$th column pair $(\tilde{\beta}_j,\tilde{\theta}_j)$, $j=1,\dots,p_{1}$, can be obtained by, 
\begin{eqnarray}\label{eq:beta_theta_tilde}
\textrm{minimize}_{(\tilde{\beta}_j,\tilde{\theta}_j)} \; v \left\{ \tilde{\beta}_j, \tilde{\theta}_j; \; \kappa_{1},\mu_{1}, 2\kappa_{1}\nu_{1}+\rho, 2\kappa_{1}\nu_{1}+\rho,\tau_{1j}^{(s)}+\rho\beta_{j}^{(s+1)}, \tau_{2j}^{(s)}+\rho\theta_{j}^{(s+1)} \right\}. 
\end{eqnarray}
Similarly, Steps 14--15 are to update $(\tilde{\boldsymbol{\zeta}},\tilde{\boldsymbol{\pi}}_{k})$, one column pair at a time, and the $k$th column pair $(\tilde{\zeta}_{k},\tilde{\pi}_{k})$, $k=1,\dots,p_{2}$, can be obtained by, 
\begin{eqnarray}\label{eq:zeta_pi_tilde}
\textrm{minimize}_{(\tilde{\zeta}_k,\tilde{\pi}_k)} \; v \left\{ \tilde{\zeta}_k,\tilde{\pi}_k; \; \kappa_{2}, \mu_{2}, 2\kappa_{2}\nu_{2}+\rho , 2\kappa_{2}\nu_{2}+\rho,\tau_{3k}^{(s)}+\rho\zeta_{k}^{(s+1)}, \tau_{4k}^{(s)}+\rho\pi_{k}^{(s+1)} \right\}. 
\end{eqnarray}
In both \eqref{eq:beta_theta_tilde} and \eqref{eq:zeta_pi_tilde}, the function $v(a_1, a_2)$ is of the form, 
\begin{equation}\label{eq:PathLasso_func}
v(a_1, a_2; \; b_1, b_2, b_3, b_4, b_5, b_6)=b_1 |a_1 a_2| + b_2 |a_1| + b_2|a_2| + \frac{1}{2}b_3 a_1^{2} + \frac{1}{2}b_4 a_2^{2} - b_5 a_1-b_6 a_2. 
\end{equation}
Its optimization has a closed-form solution; see \citet[Lemma 3.2]{zhao2016pathway}. 

Our method involves a number of tuning parameters. For $\nu_{1}$, $\nu_{2}$ and $\rho$, our simulations have found that the final estimators are not overly sensitive to their values. The same phenomenon was observed in \citet{zhao2016pathway}. We thus fix $\nu_{1}=\nu_{2}=2$ and $\rho=1$. For $(\kappa_{1},\kappa_{2},\kappa_{3},\kappa_{4},\mu_{1},\mu_{2})$, for simplicity, we set $\kappa_{1}=\kappa_{2}=\kappa_{3}=\kappa_{4}=\tilde\kappa$, and $\mu_{1}=\mu_{2}=\tilde{\mu}$. We further fix the ratio between $\tilde\kappa$ and $\tilde\mu$, following a similar tuning strategy as elastic net~\citep{zou2005regularization}. We then run a grid search to minimize a modified Bayesian information criterion (BIC), 
\begin{equation} \label{eqn:bic}
\mathrm{BIC} = -2\log L\left( \hat{\boldsymbol{\beta}},\hat{\boldsymbol{\theta}},\hat{\boldsymbol{\zeta}},\hat{\boldsymbol{\pi}},\hat{\boldsymbol{\Lambda}},\hat{\delta} \right) + \log(n)\left(|\hat{\mathcal{A}}_{1}|+|\hat{\mathcal{A}}_{2}|+|\hat{\mathcal{A}}_{3}|\right),
\end{equation}
where the estimators are obtained under a given set of tuning parameters, $\hat{\mathcal{A}}_{1} = \{j : \hat\beta_{j}\hat\theta_{j}\neq 0\}$, $\hat{\mathcal{A}}_{2} = \{k : \hat\zeta_{k}\hat\pi_{k} \neq 0\}$, and $\hat{\mathcal{A}}_{3}=\{(j,k):\hat\beta_{j}\hat\lambda_{jk}\hat\pi_{k}\neq 0\}$.

\section{Theory}
\label{sec:asmp}

In this section, we study the asymptotic properties of our proposed estimator. We consider two combinations of penalties. One is as in \eqref{eq:objfunc} with all three penalties, $P_{1}$, $P_{2}$ and $P_{3}$. The second only involves $P_{1}$ and $P_{3}$, since these two penalties alone would have achieved the selection of pathways between $X$ and $Y$. We show that the two combinations can achieve the same convergence rate in estimating the pathway effects as $p_{1},p_{2}$ and $n$ tending infinity.  To simplify the problem and focus on the mediation pathways, we assume the direct effect $\delta$ is known, and let $V=Y-X\delta$. Let $\boldsymbol{\Theta}^{*} = (\boldsymbol{\beta}^{*},\boldsymbol{\theta}^{*},\boldsymbol{\zeta}^{*},\boldsymbol{\pi}^{*},\boldsymbol{\Lambda}^{*})$ denote the true parameters, and $\hat{\boldsymbol{\Theta}} = (\hat{\boldsymbol{\beta}},\hat{\boldsymbol{\theta}},\hat{\boldsymbol{\zeta}},\hat{\boldsymbol{\pi}},\hat{\boldsymbol{\Lambda}})$ the global minimizer of the optimization in \eqref{eq:objfunc}. Let $\boldsymbol{\varsigma}^{*}=\boldsymbol{\Lambda}^{*}\boldsymbol{\pi}^{*}\in\mathbb{R}^{p_{1}}$. Let $\mathcal{S}_{1}=\{j:\beta_{j}^{*}\neq 0\}$, $\mathcal{S}_{2}=\{j:\theta_{j}^{*}\neq 0\}$, $\mathcal{S}_{3}=\{k:\zeta_{k}^{*}\neq 0\}$, $\mathcal{S}_{4}=\{k:\pi_{k}^{*}\neq 0\}$, and $\mathcal{S}_{5}=\{j:\varsigma_{j}^{*}\neq 0\}$ denote the support of $\boldsymbol{\beta}^{*}$, $\boldsymbol{\theta}^{*}$, $\boldsymbol{\zeta}^{*}$, $\boldsymbol{\pi}^{*}$ and $\boldsymbol{\varsigma}^{*}$, respectively, and $s_l = | \mathcal{S}_l |$ the cardinality of set $\mathcal{S}_l$, $l = 1, \ldots, 5$. We first introduce a set of regularity conditions. 

\begin{enumerate}[({C}1)]
\item The distribution of $X$ has a finite variance, and $|X_i| \leq c_0$ almost surely, $i=1,\dots,n$. 

\item The true parameters are bounded, in that, $|\theta_{j}^{*}|\leq c_{01}$ for $j = 1, \ldots, p_1$, $|\pi_{k}^{*}|\leq c_{02}$ for $k = 1, \ldots, p_2$, and $|\varsigma_{j}^{*}| \leq c_{03}$ for $j \in 1, \ldots, p_1$.

\item The penalty functions, evaluated at the true parameters, are bounded. That is,
\begin{enumerate}[({C3}-1)]
\item For $P_1$, $\sum_{j=1}^{p_{1}}\left\{ |\beta_{j}^{*}\theta_{j}^{*}|+\nu_{1}(\beta_{j}^{*2}+\theta_{j}^{*2}) \right\} \leq c_{11}$, and $\sum_{k=1}^{p_{2}}\left\{ |\zeta_{k}^{*}\pi_{k}^{*}|+\nu_{2}(\zeta_{k}^{*2}+\pi_{k}^{*2}) \right\}$ $\leq c_{12}$;
\item For $P_2$, $\sum_{j=1}^{p_{1}}\left(|\beta_{j}^{*}|+|\theta_{j}^{*}|\right)\leq c_{21}$, and $\sum_{k=1}^{p_{2}}\left(|\zeta_{k}^{*}|+|\pi_{k}^{*}|\right)\leq c_{22}$; 
\item For $P_3$, $\sum_{j=1}^{p_{1}}\sum_{k=1}^{p_{2}}|\lambda_{jk}^{*}|\leq c_{3}$.
\end{enumerate}

\item All the entries of the error variance terms are bounded by $c_4$.
\end{enumerate}
Condition (C1) is a standard regularity condition on the design matrix in high-dimensional regression settings. When $X$ is binary or categorical, (C1) is satisfied. When considering all three penalties $P_{1}$, $P_{2}$ and $P_{3}$, we do not actually need (C2), as (C3-2) is sufficient. But if we only consider $P_{1}$, and $P_{3}$, we impose (C2) that puts bounds on the true parameters. (C3-2) also implicitly regulates the sparsity in $\boldsymbol{\beta}^{*}$, $\boldsymbol{\theta}^{*}$, $\boldsymbol{\zeta}^{*}$ and $\boldsymbol{\pi}^{*}$. (C4) is the finite variance condition on the model errors, which is again common in the literature.

We evaluate the accuracy of our estimator by the mean squared prediction error,
\begin{equation}
\mathrm{MSPE}=\frac{1}{n}\sum_{i=1}^{n}\left(\hat{V}_{i}-V_{i}^{*}\right)^{2},
\end{equation}
where $\hat{V}_{i}$ and $V_{i}^{*}$ are the predicted pathway effects under the estimated parameter $\hat{\boldsymbol{\Theta}}$ and the true parameter $\boldsymbol{\Theta}^{*}$, respectively, for subject $i$, $i=1,\dots,n$. The predicted pathway effects under our mediation model setting are defined as follows. 
\begin{definition}\label{def:prediction}
For a treatment condition $X=x$, define
\begin{enumerate}[(1)]
\item The predicted outcome through $\bM_{1}$, but not through $\bM_{2}$, is: 
$\hat{V}_{1}=x\hat{\boldsymbol{\beta}}\hat{\boldsymbol{\theta}}=x\left(\sum_{j=1}^{p_{1}}\hat{\beta}_{j}\hat{\theta}_{j}\right)$.

\item The predicted outcome through $\bM_{2}$, but not through $\bM_{1}$, is: 
$\hat{V}_{2}=x\hat{\boldsymbol{\zeta}}\hat{\boldsymbol{\pi}}=x\left(\sum_{k=1}^{p_{2}}\hat{\zeta}_{k}\hat{\pi}_{k}\right)$. 

\item The prediction through both $\bM_{1}$ and $\bM_{2}$ is: 
$\hat{V}_{3}=x\hat{\boldsymbol{\beta}}\hat{\boldsymbol{\Lambda}}\hat{\boldsymbol{\pi}}=x\left(\sum_{j=1}^{p_{1}}\sum_{k=1}^{p_{2}}\hat{\beta}_{j}\hat{\lambda}_{jk}\hat{\pi}_{k}\right)$. 

\item The total prediction of $V$ is: $\hat{V}=\hat{V}_{1}+\hat{V}_{2}+\hat{V}_{3}$. 
\end{enumerate}
\end{definition}

\begin{theorem}\label{thm:MSIPE_est}
We have the following results regarding the pathway effect estimation. 
\begin{enumerate}[(i)]
\item Under penalties $P_{1}$ and $P_{3}$, suppose (C1), (C2), (C3-1), (C3-3) and (C4) hold, then
\begin{eqnarray*}\label{eq:eMSIPE_P1P3}
\mathbb{E}(\mathrm{MSPE}) & \leq &   
2 c_0 c_4 \left\{ c_{11} s_{2} c_{01} \sqrt{\frac{2\log(2p_{1})}{n}} + c_{12} s_{4} c_{02} \sqrt{\frac{2\log(2p_{2})}{n}} \right. \\
& & \quad\quad\quad \left. + \; c_3 \sqrt{c_{11} c_{12} / \nu_1 \nu_2} (1+s_{5} c_{03})\sqrt{\frac{2\log(2p_{1})}{n}} \right\}, \\
\|\hat{\boldsymbol{\beta}}\hat{\boldsymbol{\theta}}-\boldsymbol{\beta}^{*}\boldsymbol{\theta}^{*}\|_{2}^{2} &\leq& \frac{1}{c_{5}}\left\{ 8c_{11}^{2}c_{0}^{2}\sqrt{\frac{2\log(2)}{n}}+2c_{0}c_{4} c_{11}s_{2}c_{01}\sqrt{\frac{2\log(2p_{1})}{n}} \right\}, \label{eq:asmp_IEM1_P1P3} \\
\|\hat{\boldsymbol{\zeta}}\hat{\boldsymbol{\pi}}-\boldsymbol{\zeta}^{*}\boldsymbol{\pi}^{*}\|_{2}^{2} &\leq& \frac{1}{c_{5}}\left\{ 8c_{12}^{2}c_{0}^{2}\sqrt{\frac{2\log(2)}{n}}+2c_{0}c_{4} c_{12}s_{4}c_{02}\sqrt{\frac{2\log(2p_{2})}{n}} \right\}, \label{eq:asmp_IEM2_P1P3} \\
\|\hat{\boldsymbol{\beta}}\hat{\boldsymbol{\Lambda}}\hat{\boldsymbol{\pi}}-\boldsymbol{\beta}^{*}\boldsymbol{\Lambda}^{*}\boldsymbol{\pi}^{*}\|_{2}^{2} &\leq& \frac{1}{c_{5}}\left\{ 8c_{3}^{2}(c_{11}c_{12}/\nu_{1}\nu_{2})c_{0}^{2}\sqrt{\frac{2\log(2)}{n}} \right. \\
&& \quad\quad \left. + \; 2c_{0}c_{4} c_{3}\sqrt{c_{11}c_{12}/\nu_{1}\nu_{2}}(1+s_{5}c_{03})\sqrt{\frac{2\log(2p_{1})}{n}} \right\}. \label{eq:asmp_IEM1M2_P1P3}
\end{eqnarray*}

\item Under penalties $P_{1}$, $P_{2}$ and $P_{3}$, suppose (C1), (C3) and (C4) hold, then
\begin{eqnarray*}\label{eq:eMSIPE_P2P3}
\mathbb{E}(\mathrm{MSPE}) & \leq &   
2 c_0 c_4 \left\{ c_{11} c_{21} \sqrt{\frac{2\log(2p_{1})}{n}} + c_{12} c_{22} \sqrt{\frac{2\log(2p_{2})}{n}} \right. \\
& & \quad\quad\quad \left. + \; c_3 \sqrt{c_{11} c_{12} / \nu_1 \nu_2} (1+c_{3}c_{22})\sqrt{\frac{2\log(2p_{1})}{n}} \right\}, \\
\|\hat{\boldsymbol{\beta}}\hat{\boldsymbol{\theta}}-\boldsymbol{\beta}^{*}\boldsymbol{\theta}^{*}\|_{2}^{2} &\leq& \frac{1}{c_{5}}\left\{ 8c_{11}^{2}c_{0}^{2}\sqrt{\frac{2\log(2)}{n}}+2c_{0}c_{4} c_{11}c_{21}\sqrt{\frac{2\log(2p_{1})}{n}}\right \}, \label{eq:asmp_IEM1_P1P2P3} \\
\|\hat{\boldsymbol{\zeta}}\hat{\boldsymbol{\pi}}-\boldsymbol{\zeta}^{*}\boldsymbol{\pi}^{*}\|_{2}^{2} &\leq& \frac{1}{c_{5}}\left\{ 8c_{12}^{2}c_{0}^{2}\sqrt{\frac{2\log(2)}{n}}+2c_{0}c_{4} c_{12}c_{22}\sqrt{\frac{2\log(2p_{2})}{n}} \right\}, \label{eq:asmp_IEM2_P1P2P3} \\
\|\hat{\boldsymbol{\beta}}\hat{\boldsymbol{\Lambda}}\hat{\boldsymbol{\pi}}-\boldsymbol{\beta}^{*}\boldsymbol{\Lambda}^{*}\boldsymbol{\pi}^{*}\|_{2}^{2} &\leq& \frac{1}{c_{5}}\left\{ 8c_{3}^{2}(c_{11}c_{12}/\nu_{1}\nu_{2})c_{0}^{2}\sqrt{\frac{2\log(2)}{n}} \right. \\
&& \quad\quad \left. + \; 2c_{0}c_{4} c_{3}\sqrt{c_{11}c_{12}/\nu_{1}\nu_{2}}(1+c_{3}c_{22})\sqrt{\frac{2\log(2p_{1})}{n}} \right\}. \label{eq:asmp_IEM1M2_P1P2P3}
\end{eqnarray*}
\end{enumerate}
\end{theorem}

\noindent
Theorem~\ref{thm:MSIPE_est} shows that, under both penalty combinations, the mean squared prediction error converges. It also shows that all three types of total pathway effects estimators are consistent in $\ell_{2}$-norm. Moreover, the convergence rate of the total indirect effect $\mathrm{IE}^{1}$ and $\mathrm{IE}^{2}$ are $\sqrt{\log(p_{1})/n}$ and $\sqrt{\log(p_{2})/n}$, respectively, which are consistent with the rate under only one set of mediators as studied in \citet{zhao2016pathway}. When considering the indirect effect through both $\bM_{1}$ and $\bM_{2}$, $\boldsymbol{\varsigma}=\boldsymbol{\Lambda}\boldsymbol{\pi}$ summarizes the post-$\bM_{1}$ pathway effects, and the problem degenerates to the case with a single set of $p_{1}$ mediators. As such, the convergence rate is $\sqrt{\log(p_{1})/n}$, and depends on the number of nonzero elements in $\boldsymbol{\varsigma}$. 

\section{Simulations}
\label{sec:sim}

In this section, we investigate the finite-sample performance of our method. We generate data following Model~\eqref{eq:LSEM_reduced}. In the interest of space, we report the generative scheme and the corresponding signal-to-noise ratio in Section~\ref{appendix:sec:sim} of the supporting information. We consider two dimension sizes $p_{1}=20$, $p_{2}=30$, with the sparsity level set at $0.1$, and $p_{1}=p_{2}=100$, with the sparsity level at $0.01$. We also consider two sample sizes $n=50$ and $n=500$. We compare different combinations of penalty functions: (i) $P_2$ and $P_3$, which is essentially a Lasso solution (P2P3); (ii) $P_1$ and $P_3$, as we discuss in Section~\ref{sec:asmp} (P1P3); and (iii) $P_1, P_2$ and $P_3$, our proposed method. For the last case, we consider two ratios between the tuning parameters, one with $\tilde{\mu} / \tilde\kappa = 1$ (P1P2P3-1) and the other with $\tilde{\mu} / \tilde\kappa = 0.1$ (P1P2P3-2). 

\begin{figure}
  \begin{center}
    \subfloat[ROC ($p_{1}=20, p_{2}=30$)]{\includegraphics[width=0.3\textwidth]{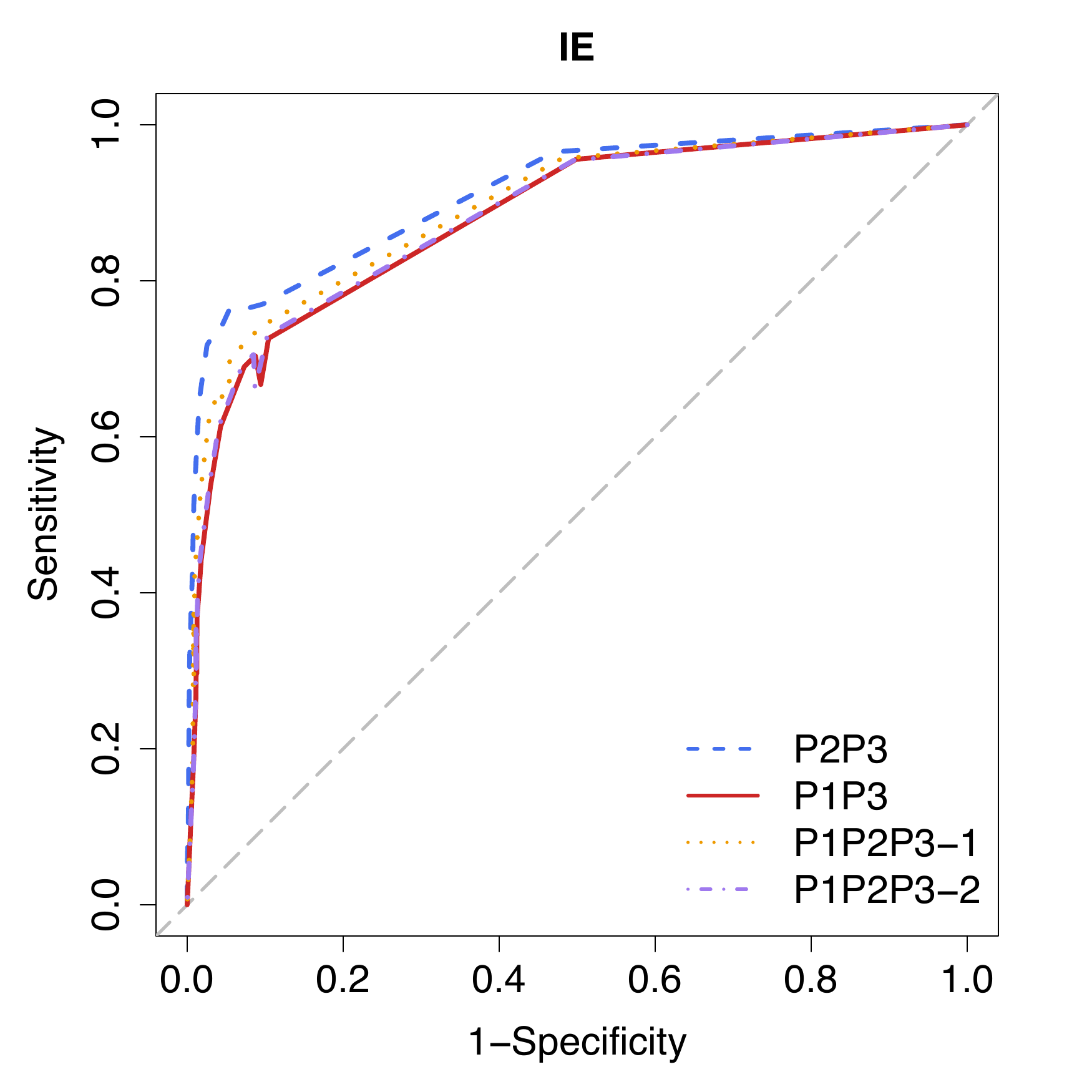}}
    \enskip{}
    \subfloat[MSE ($p_{1}=20, p_{2}=30$)]{\includegraphics[width=0.3\textwidth]{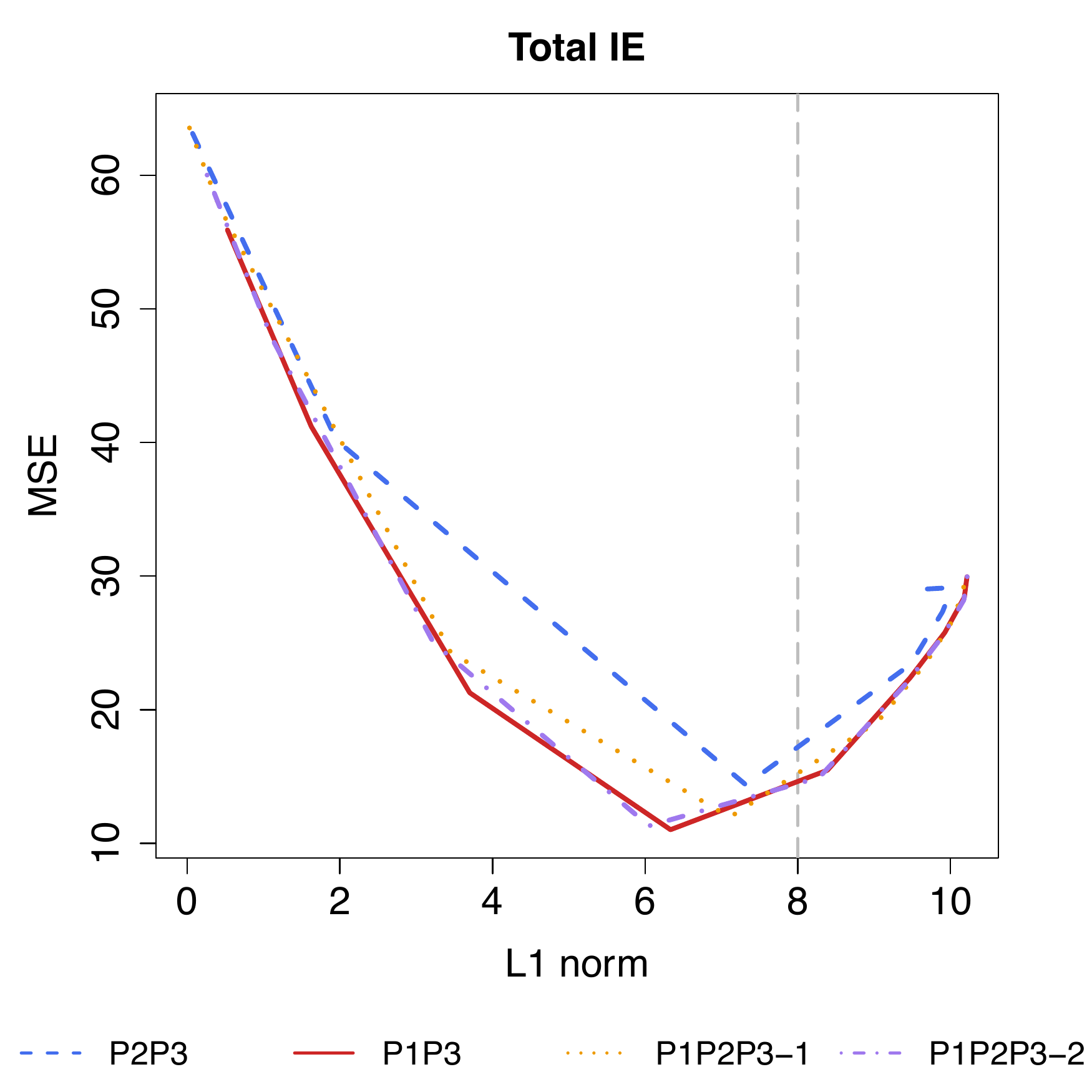}}
    \enskip{}
    \subfloat[Time ($p_{1}=20, p_{2}=30$)]{\includegraphics[width=0.3\textwidth]{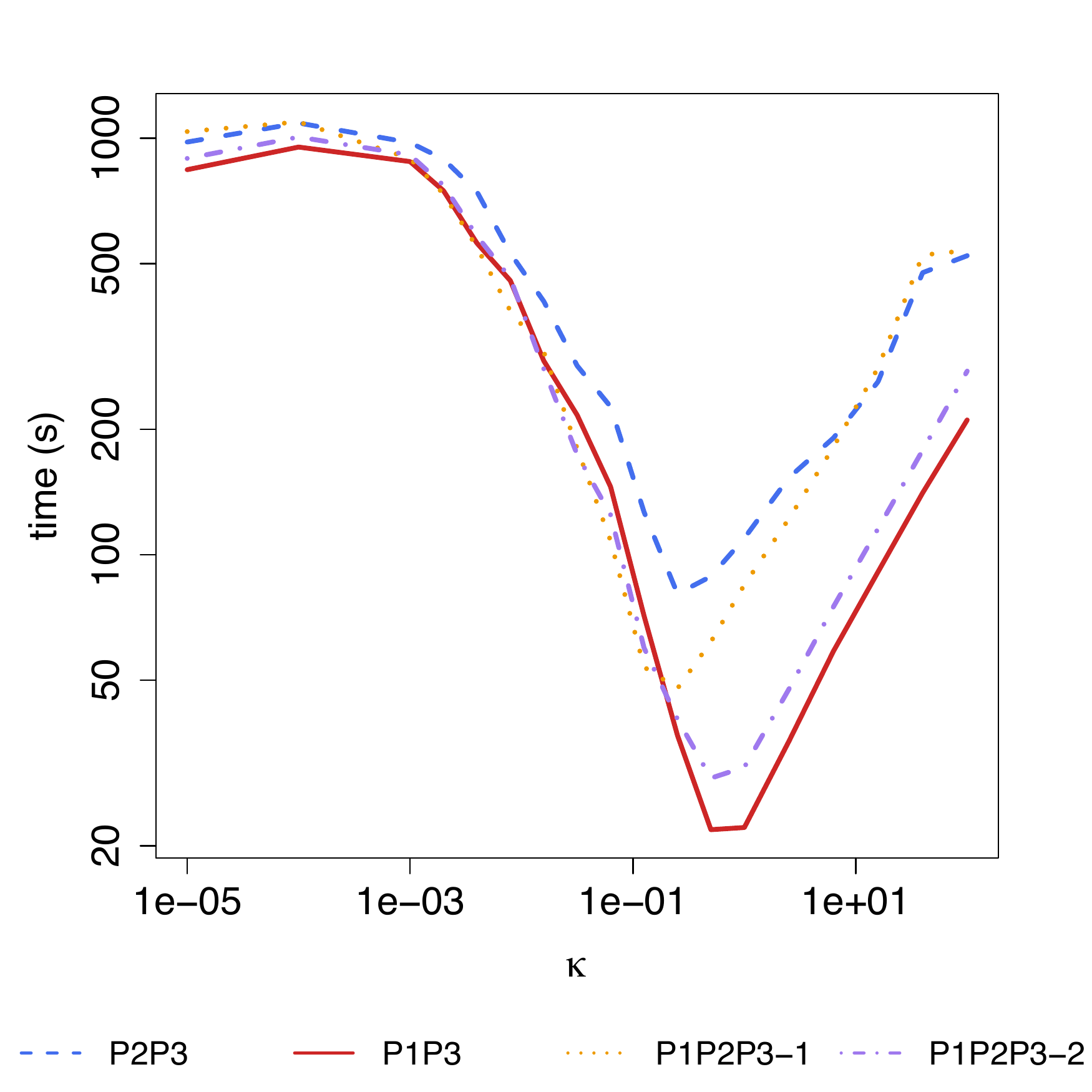}}
            
    \subfloat[ROC ($p_{1}=p_{2}=100$)]{\includegraphics[width=0.3\textwidth]{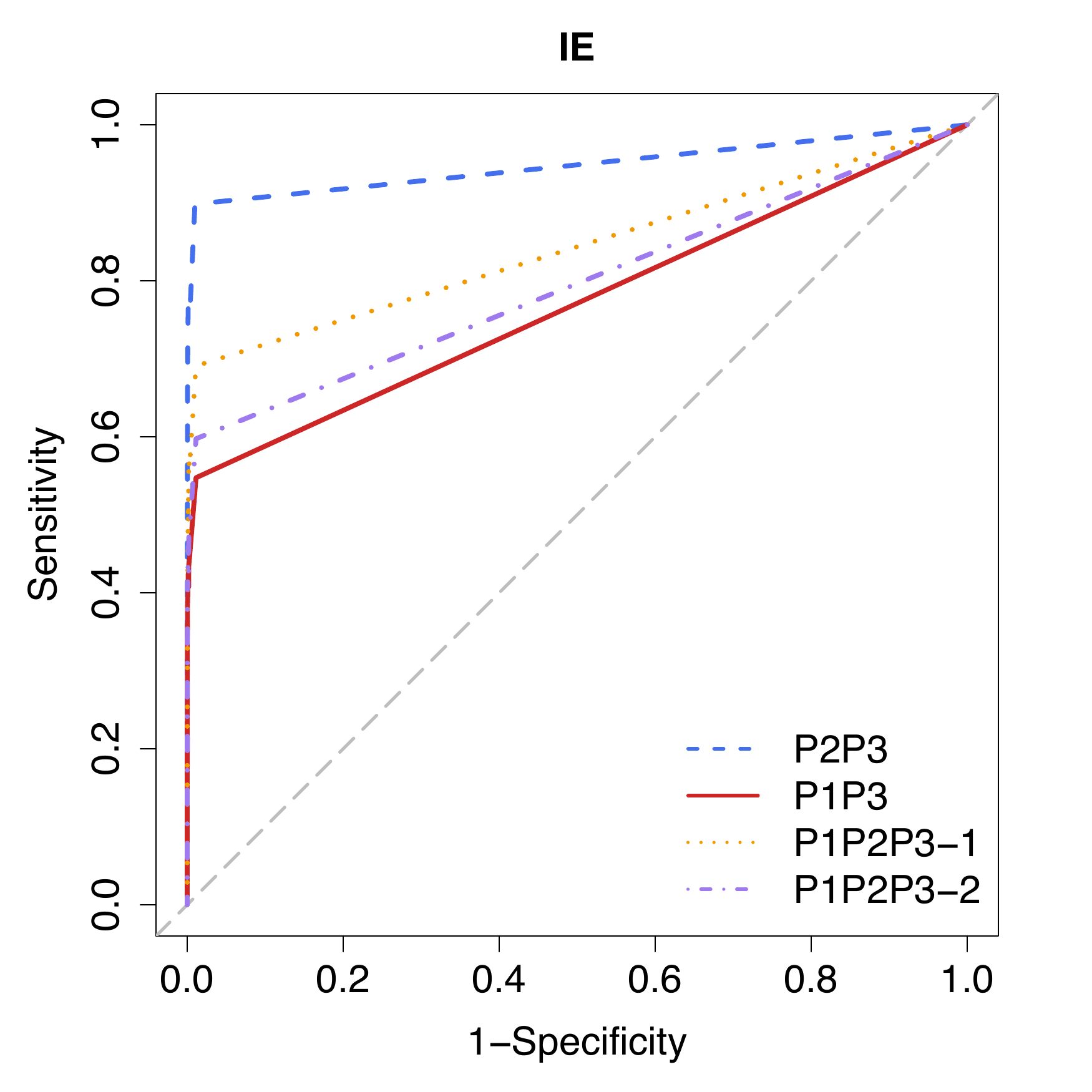}}
    \enskip{}
    \subfloat[MSE ($p_{1}=p_{2}=100$)]{\includegraphics[width=0.3\textwidth]{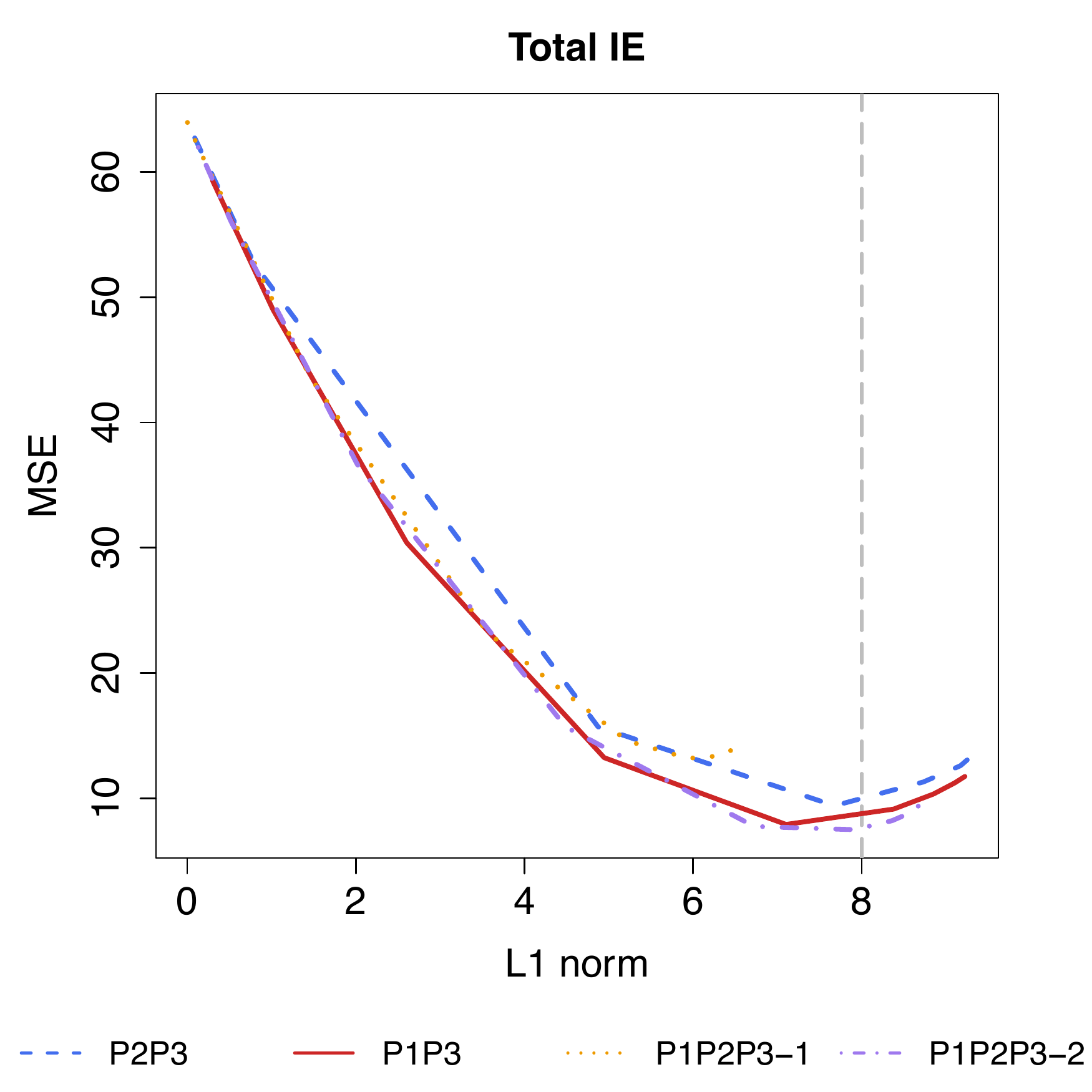}}
    \enskip{}
    \subfloat[Time ($p_{1}=p_{2}=100$)]{\includegraphics[width=0.3\textwidth]{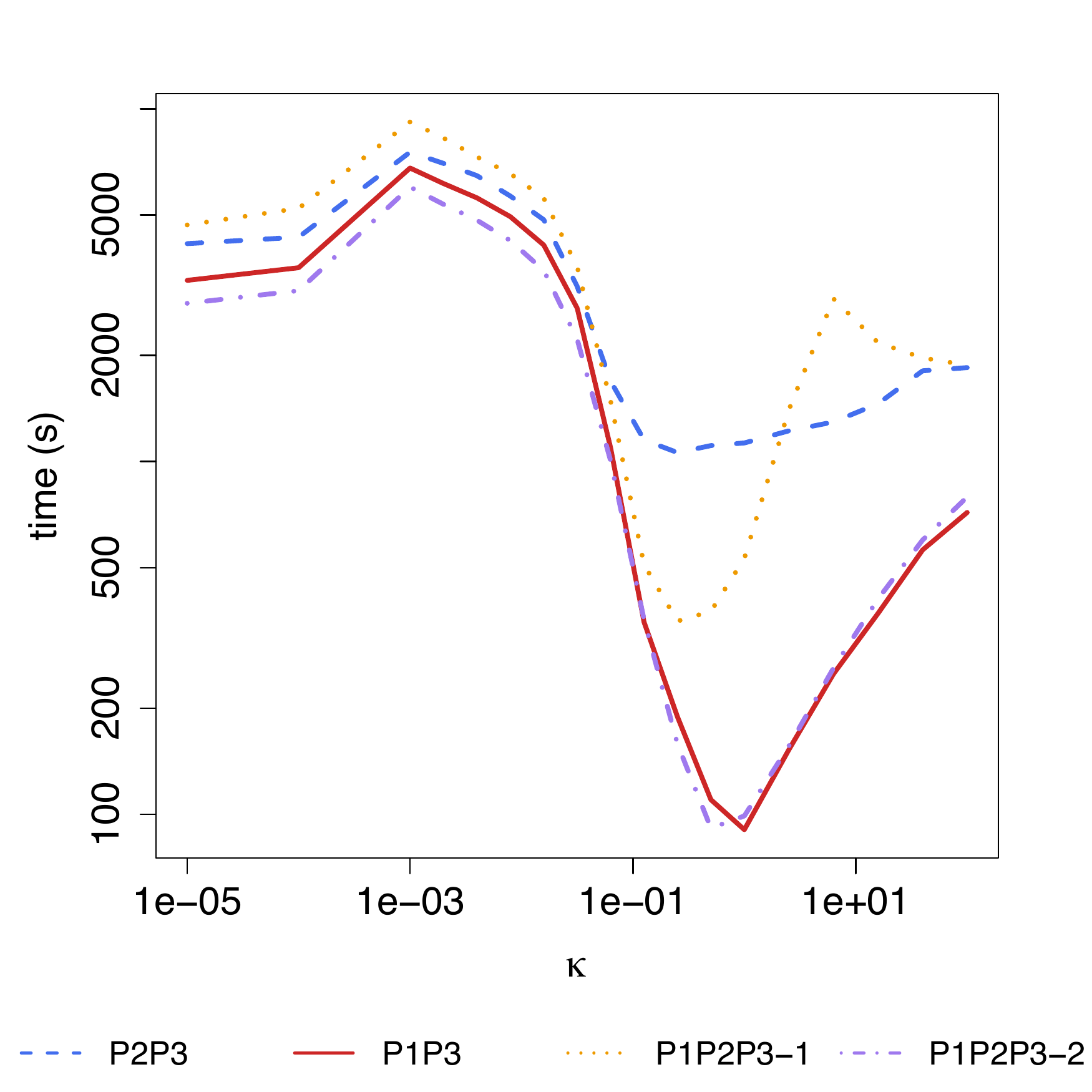}}
  \end{center}
  \caption{\label{fig:sim-n50} The ROC curve of the indirect effect pathway selection, the MSE of the total indirect effect, and the computation time, as functions of the tuning parameter $\tilde\kappa$. The sample size is $n=50$.}
\end{figure}

Figure~\ref{fig:sim-n50} reports the average simulation results based on 200 data replications with $n=50$, as the tuning parameter $\tilde\kappa$ varies. The evaluation criteria include the receiver operating characteristic (ROC) curve of the identification of all the nonzero indirect effects, the mean squared error (MSE) of estimating the total indirect effect IE as defined in \eqref{eqn:total-effect-decomp}, and the corresponding computation time in seconds. Table~\ref{tab:sim-n50} reports the results with the tuning parameter $\tilde\kappa$ selected by the BIC criterion in \eqref{eqn:bic}. The evaluation criteria include the sensitivity, specificity, MSE, and the true and actual estimate of the total indirect effect. We also report the results with $n=500$ in Section~\ref{appendix:sec:sim} of the supporting information. From these figures and tables, we see that there is a trade-off between the estimation accuracy and the selection accuracy. The method with all three penalties (P1P2P3-2) achieves a competitive overall performance.  

\begin{table}
\caption{\label{tab:sim-n50}The estimate and the mean squared error of the total indirect effect, and the sensitivity and specificity of the indirect effect pathway selection, with the tuning parameter $\tilde\kappa$ selected by BIC. The sample size is $n=50$.}
\begin{center}
\begin{tabular}{l l R{2cm} R{2cm} R{2cm} R{2cm}} 
\hline
Mediator dimension & Criterion & \multicolumn{1}{c}{P2P3} & \multicolumn{1}{c}{P1P3} & \multicolumn{1}{c}{P1P2P3-1} & \multicolumn{1}{c}{P1P2P3-2} \\ 
    \hline
    & Truth & \multicolumn{4}{c}{8} \\
    & Estimate & 8.190 & 8.122 & 8.208 & 8.148 \\
    & MSE & 29.67 & 31.35 & 29.70 & 31.01 \\
    & Sensitivity & 0.620 & 0.539 & 0.560 & 0.541 \\
    \multirow{-5}{*}{$p_{1}=20,=p_{2}=30$} & Specificity & 0.973 & 0.965 & 0.972 & 0.966 \\ 
    \hline
    & Truth & \multicolumn{4}{c}{8} \\
    & Estimate & 9.445 & 9.503 & 7.797 & 9.169 \\
    & MSE & 13.72 & 13.52 & 12.06 & 11.69 \\
    & Sensitivity & 0.721 & 0.417 & 0.517 & 0.439 \\
    \multirow{-5}{*}{$p_{1}=p_{2}=100$} & Specificity & 0.999 & 0.999 & 0.999 & 0.999 \\ \hline
\end{tabular}
\end{center}
\end{table}

\section{Data Analysis}
\label{sec:real}

We revisit our motivation example in Section~\ref{sec:introduction}. We analyzed a set of $n=136$ participants from the recent S1200 release of the Human Connectome Project. The outcome of interest is a language behavior measure, the picture vocabulary test, which evaluates language and vocabulary comprehension. In the test, an audio recording of a word and four photographic images were presented to the participants on a screen, who responded by choosing the image that most closely matches the meaning of the word. Statistically significant sex difference was observed in this test after age adjustment under a linear model. The two sets of mediators are DTI and resting-state fMRI measures. The DTI images were preprocessed following the pipeline of \cite{zhang2019tensor}. From each DTI scan, we obtained a 
symmetric structural connectivity matrix, with nodes corresponding to the brain regions-of-interest based on the Desikan Atlas \citep{desikan2006automated}, and the edges recording the number of white fiber pathways \citep{zhang2019tensor}, a measure of structural connectivity between pairs of regions. We then removed the regions with zero connectivity in over 25\% subjects, vectorized the upper triangular connectivity matrix, and obtained a $p_1=531$-dimensional vector of DTI measures. The fMRI images were preprocessed following the pipeline of \cite{glasser2013minimal}. From each fMRI scan, we obtained a 
symmetric functional connectivity matrix, with nodes corresponding to the brain regions based on the Harvard-Oxford Atlas of FSL \citep{smith2004advances}, and the edges recording the $z$-transformed Pearson correlation. We then focused on the brain regions corresponding to those of DTI, vectorized the upper triangular connectivity matrix, and obtained a $p_2=917$-dimensional vector of fMRI measures. 

\begin{figure}
\begin{center}
\includegraphics[width=\textwidth]{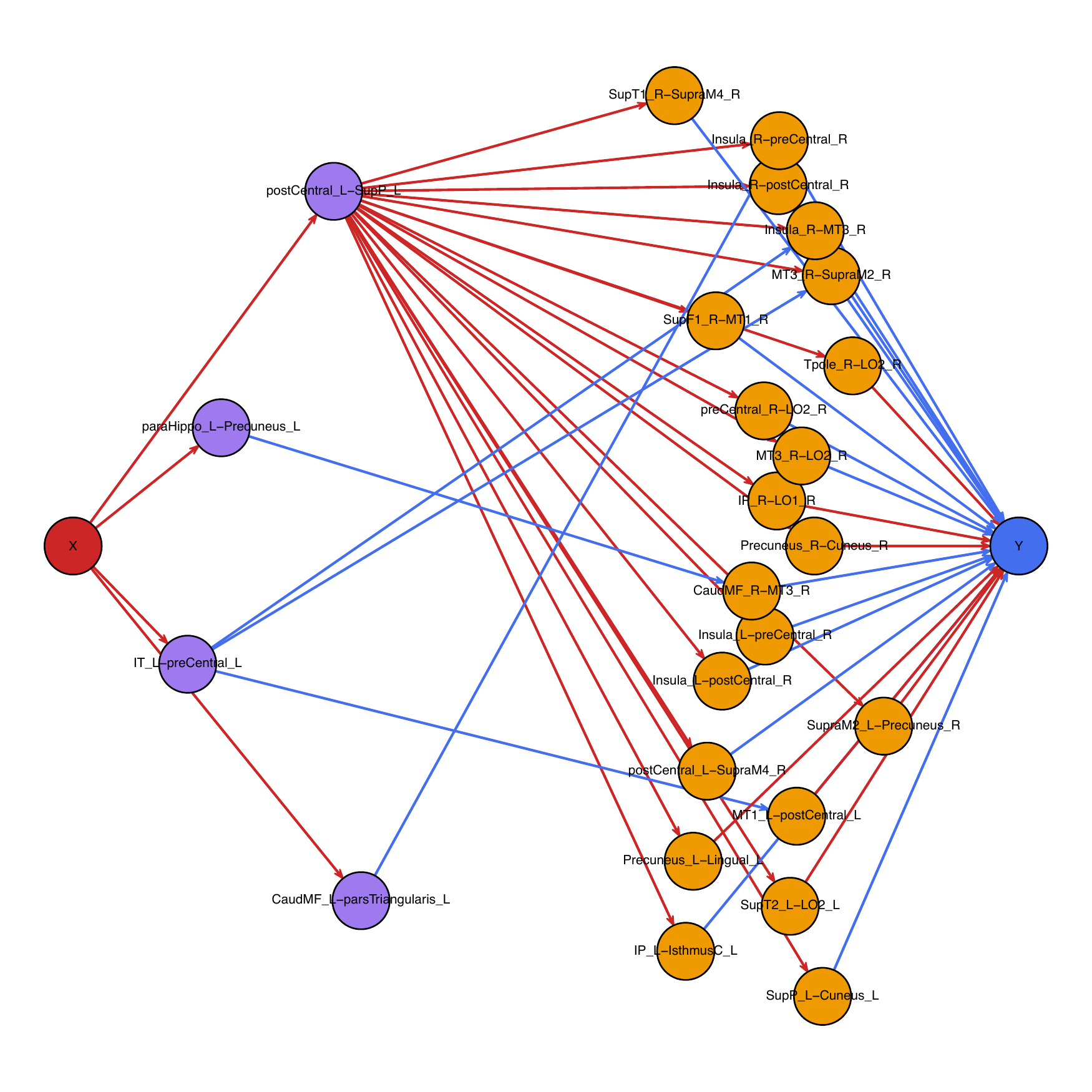}
\end{center}
\caption{\label{fig:real_path} The estimated pathways with picture vocabulary test performance as the outcome ($Y$) when comparing male ($X=1$) versus female ($X=0$). The nodes in purple are from brain structural connectivity, and nodes in orange are from brain functional connectivity. The edges in red indicate positive effects, and the ones in blue indicate negative effects.}
\end{figure}

We applied the proposed method to this data. Since P1P2P3-2 achieved the overall best performance in both selection and estimation accuracy in simulations, we employed this penalty combination. We observed a significant sex difference total effect, with a $p$-value of 0.016. This total effect can be decomposed following \eqref{eqn:total-effect-decomp}. Specifically, the penalized estimate of direct effect was zero, suggesting that the difference can be fully explained by the variations in brain connectivity. The estimated total indirect effect due to  the structural connectivity alone, the functional connectivity alone, and both connectivities was 3.322, 0.297, and -0.109, respectively. Figure~\ref{fig:real_path} presents the identified brain pathways through both the structural and functional connectivities. 

\begin{figure}
\begin{center}
\subfloat[]{{\includegraphics[width=0.25\textwidth,trim={0cm 0.5cm 8.5cm 0},clip]{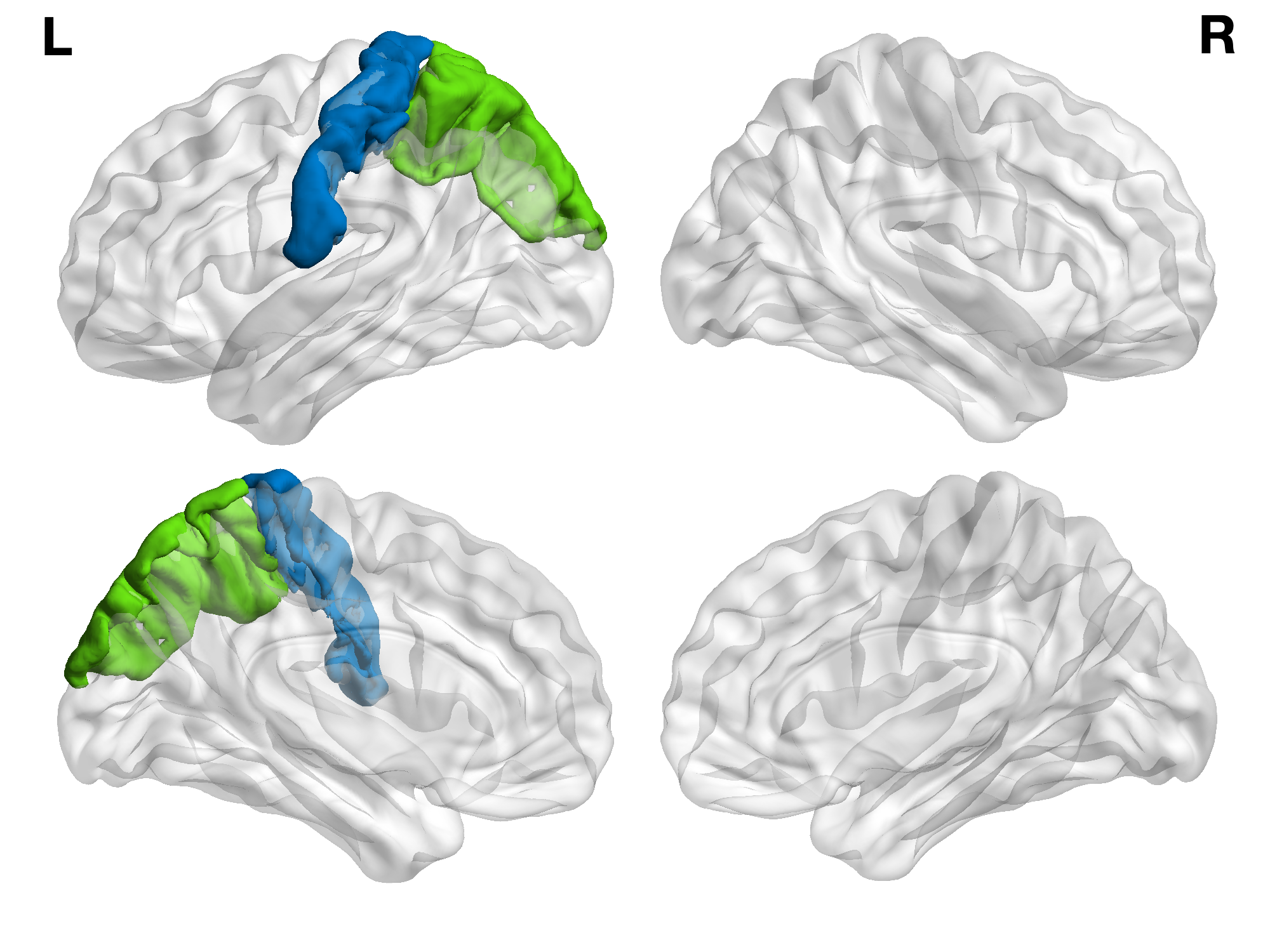}}}
\enskip
\subfloat[]{{\includegraphics[width=0.25\textwidth,trim={0cm 0.5cm 8.5cm 0},clip]{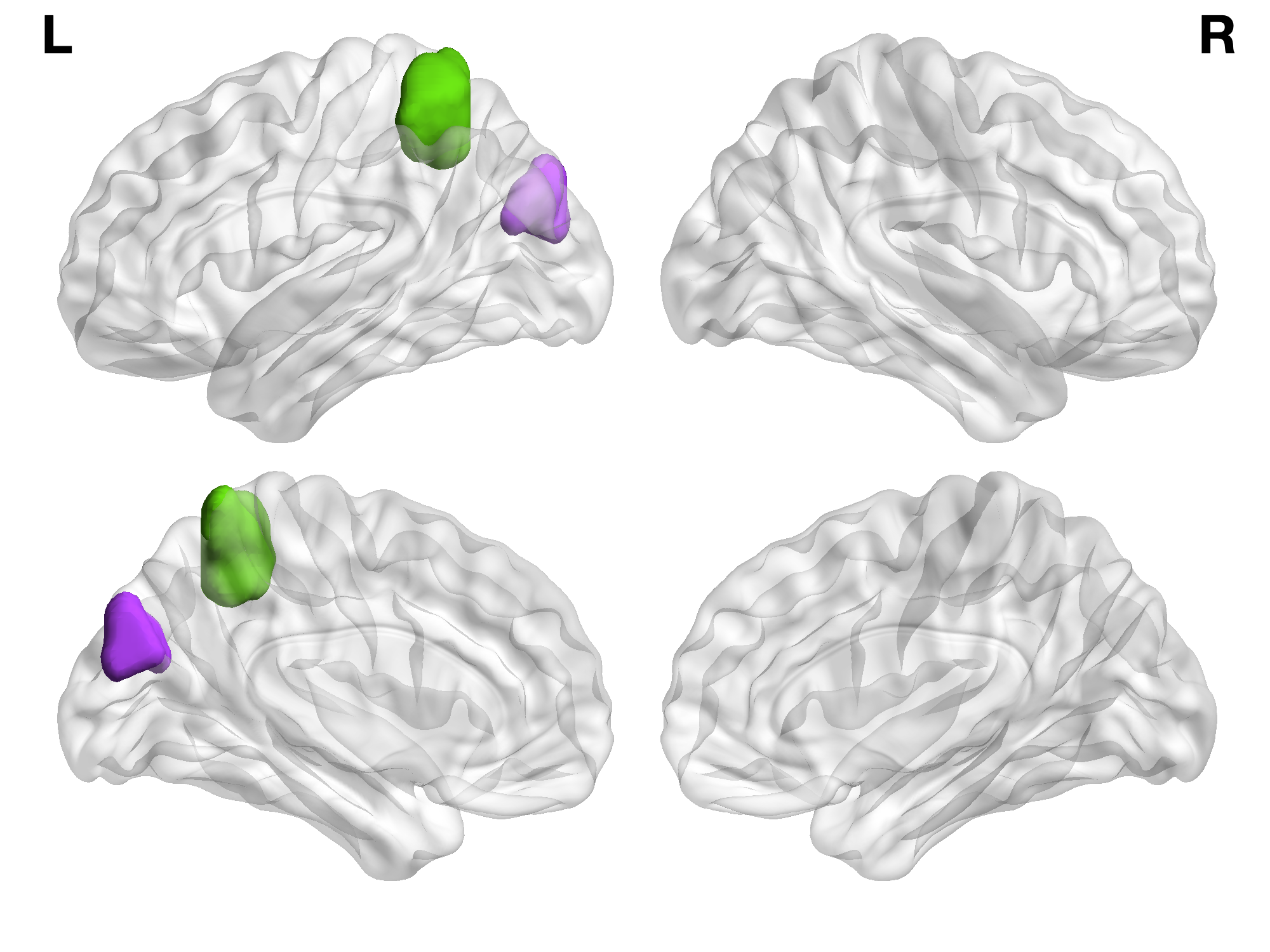}}}
\enskip
\subfloat[]{{\includegraphics[width=0.25\textwidth,trim={0cm 0.5cm 8.5cm 0},clip]{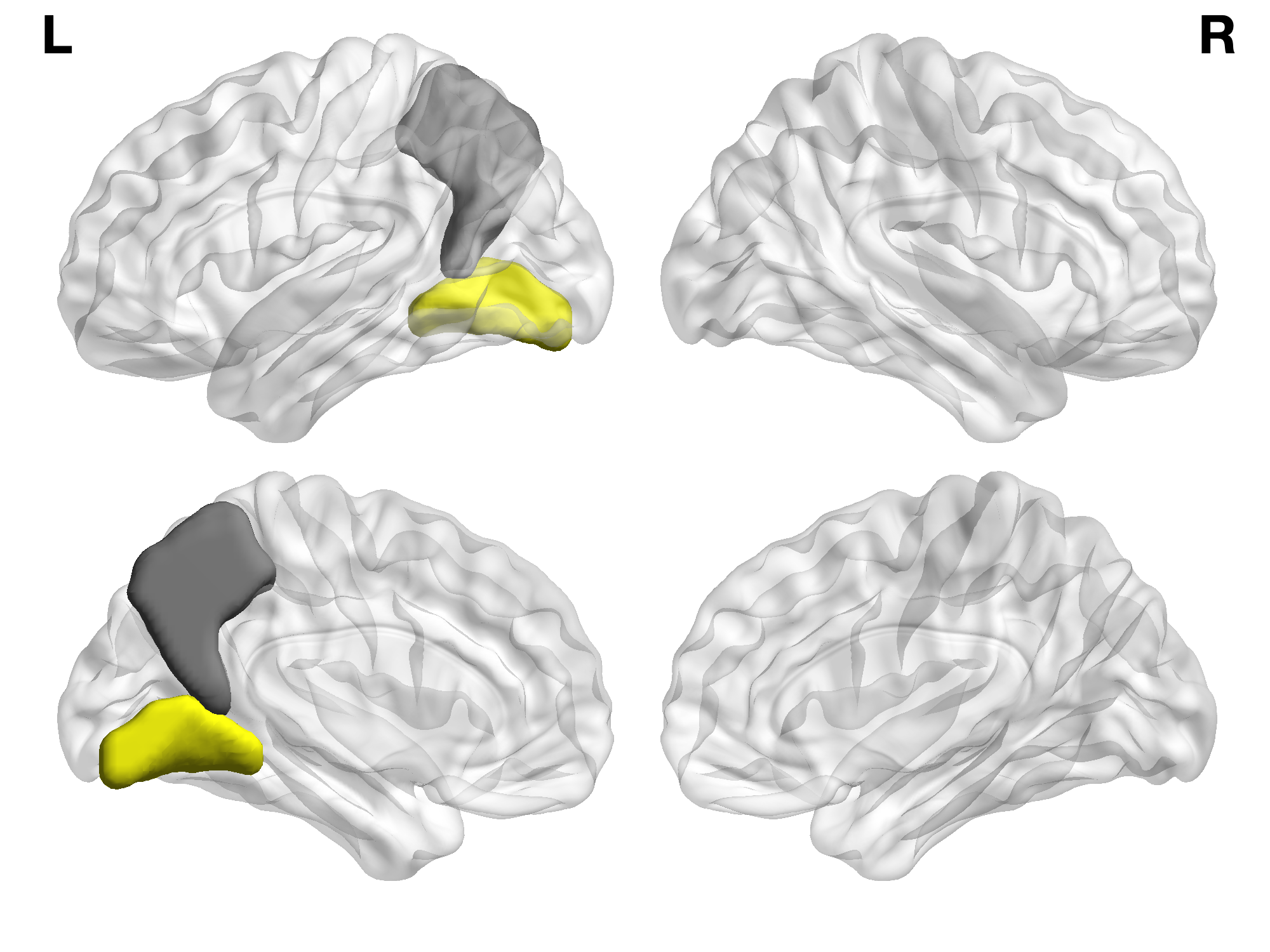}}}
\enskip
\includegraphics[width=0.15\textwidth]{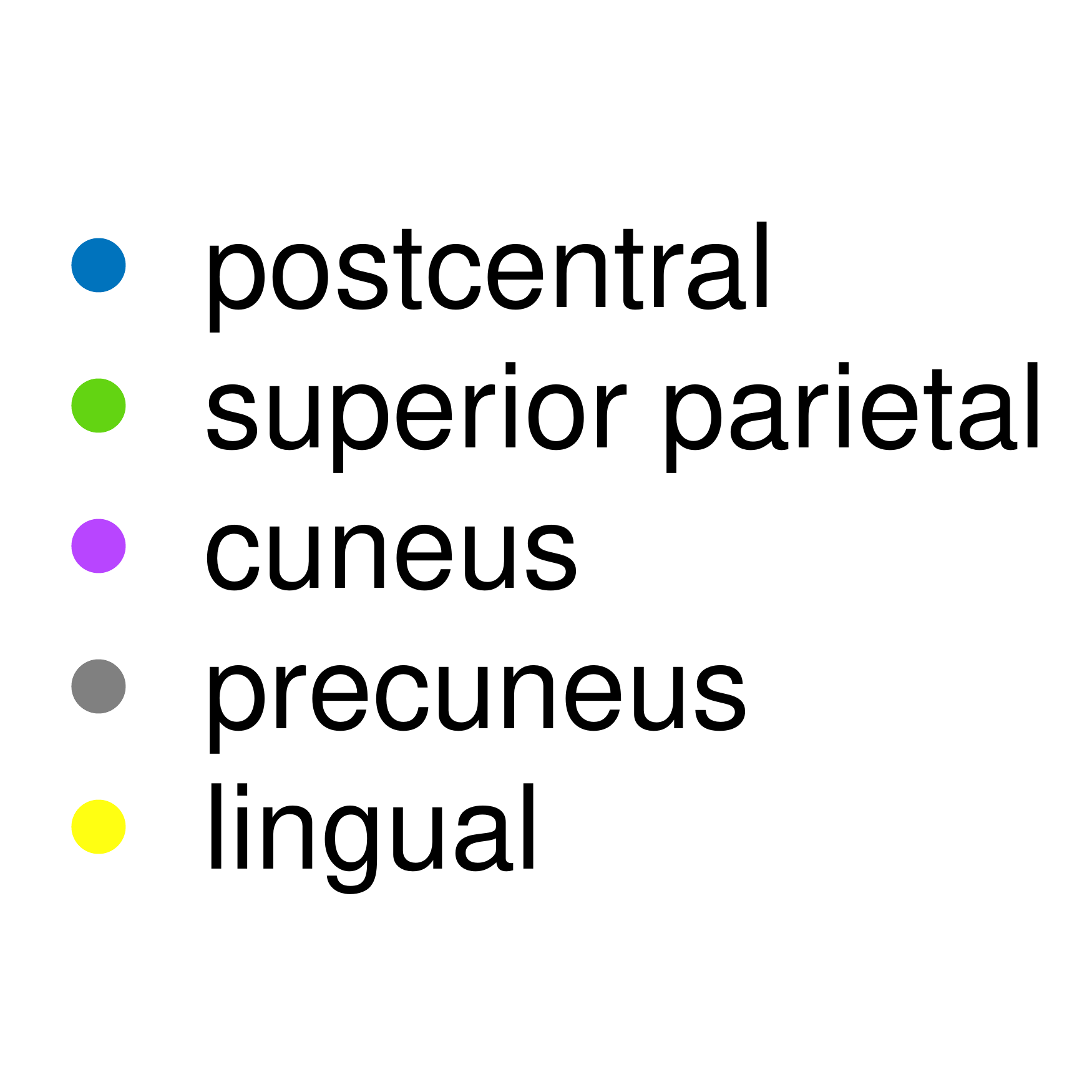}
\end{center}
\caption{\label{fig:real_brain_memory}The identified brain pathway related to working memory. (a) DTI: postcentral gyrus (Left) -- superior parietal lobule (Left), (b) fMRI: superior parietal lobule (Left) -- cuneus (Left), (c) fMRI: precuneus (Left) -- lingual gyrus (Left).}
\end{figure}

Among these pathways, one group involves structural connectivity between left postcentral gyrus (\texttt{postCentral\_L}) and left superior parietal lobule (\texttt{SupP\_L}), then via functional connectivity between numerous brain regions. Figure~\ref{fig:real_brain_memory} shows some of these pathways through these structural connections and two functional connectivities, between left superior parietal lobule and left cuneus (\texttt{Cuneus\_L}), and between left precuneus (\texttt{Precuneus\_L}) and left lingual gyrus (\texttt{Lingual\_L}). These pathways are suggestive of working memory pathways. Postcentral gyrus and cuneus have been identified in visual processing, and the cuneus as a mid-level visual processing area has been found to be modulated by working memory~\citep{salmon1996regional}. Precuneus, as part of the default mode network, is involved in working memory, especially for tasks related to verbal processing~\citep{wallentin2006parallel}. Lingual gyrus, located in the occipital lobe, plays an important role in visual processing. Left lingual gyrus is found activated during memorization~\citep{kozlovskiy2014activation}, and is involved in tasks related to naming and word recognition~\citep{mechelli2000differential}. 

\begin{figure}
\begin{center}
\subfloat[]{{\includegraphics[width=0.3\textwidth,trim={1cm 1cm 1cm 1cm},clip]{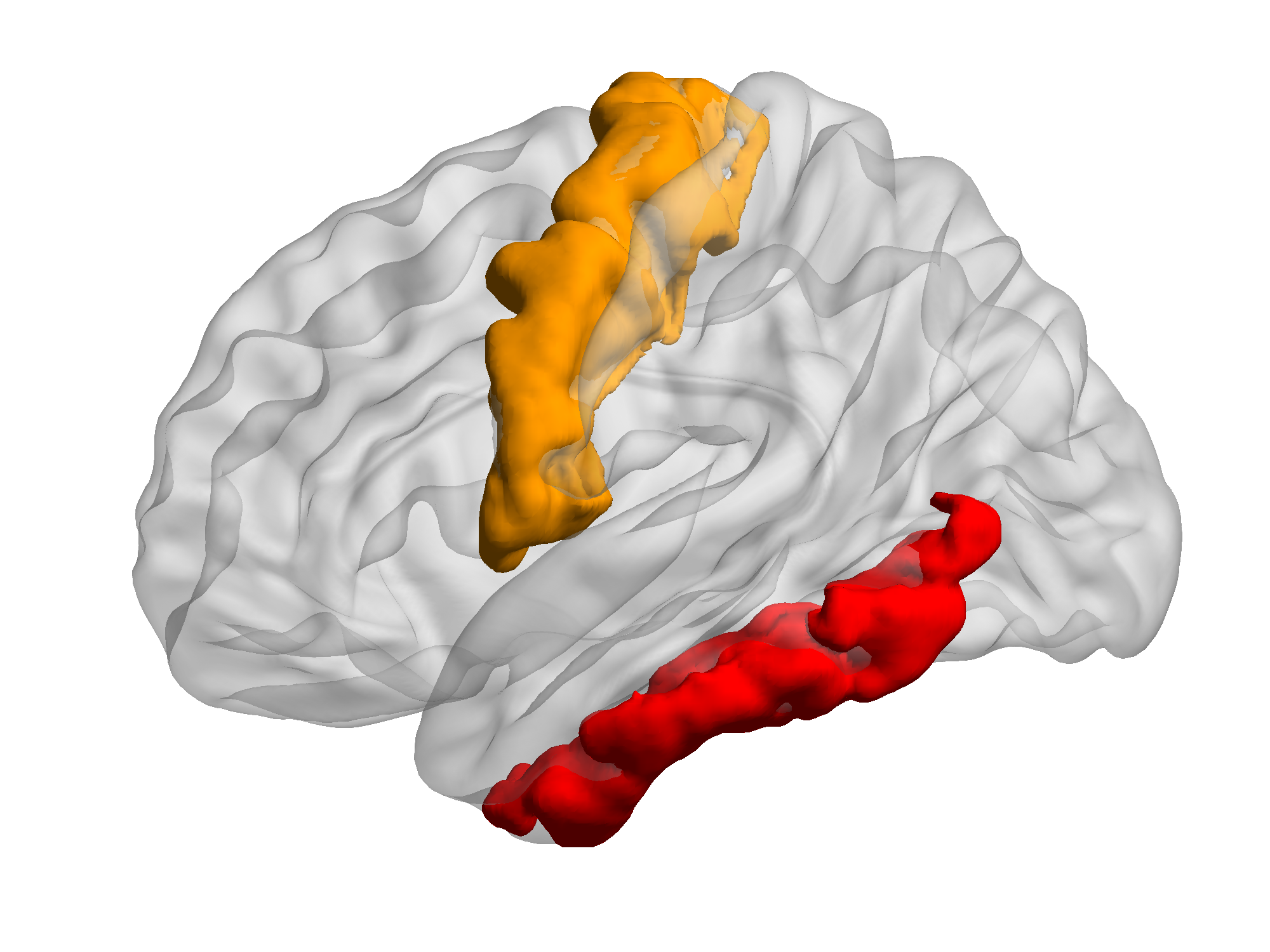}}}
\enskip
\subfloat[]{{\includegraphics[width=0.3\textwidth,trim={1cm 1cm 1cm 1cm},clip]{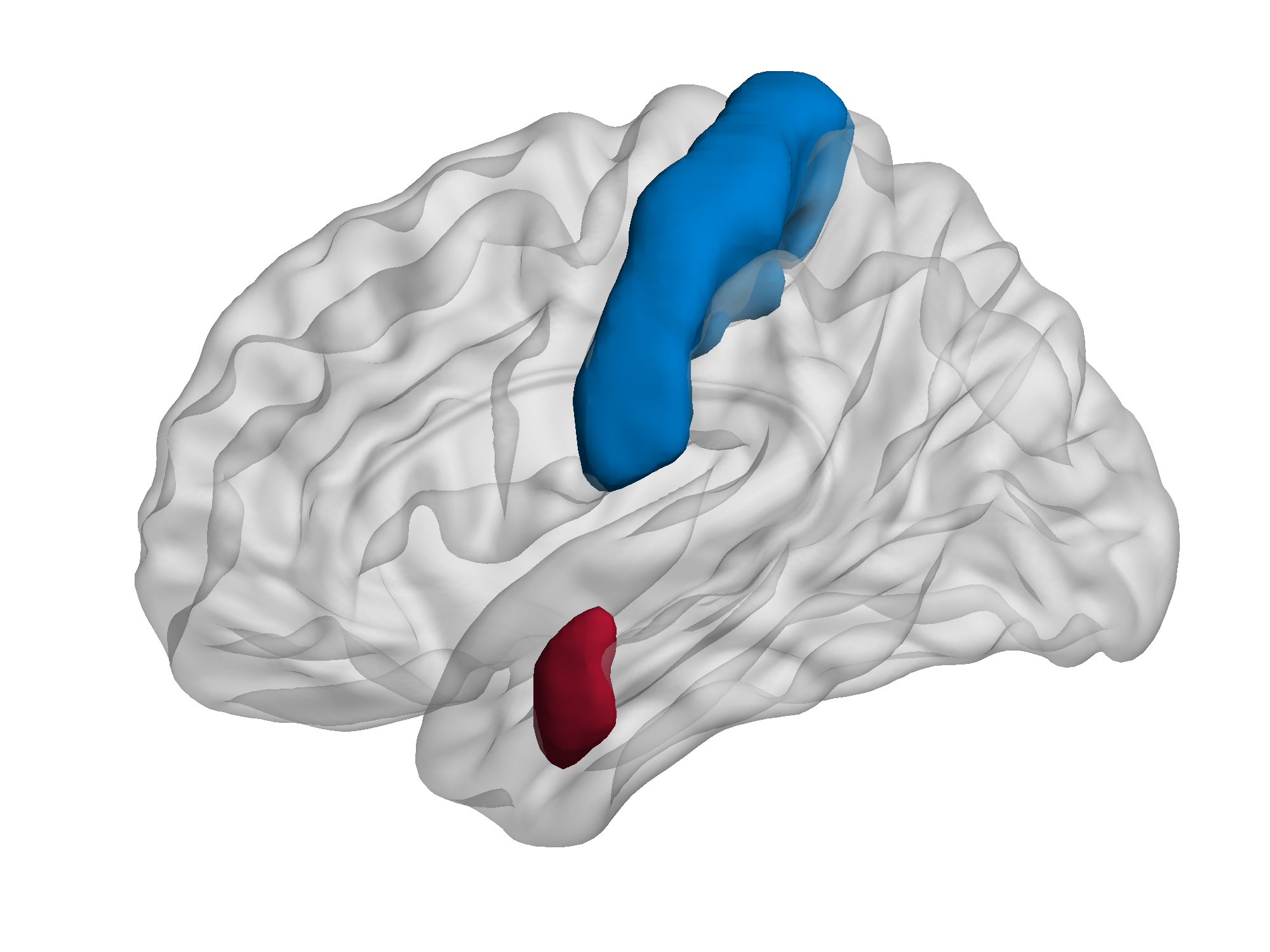}}}
\enskip
\includegraphics[width=0.15\textwidth]{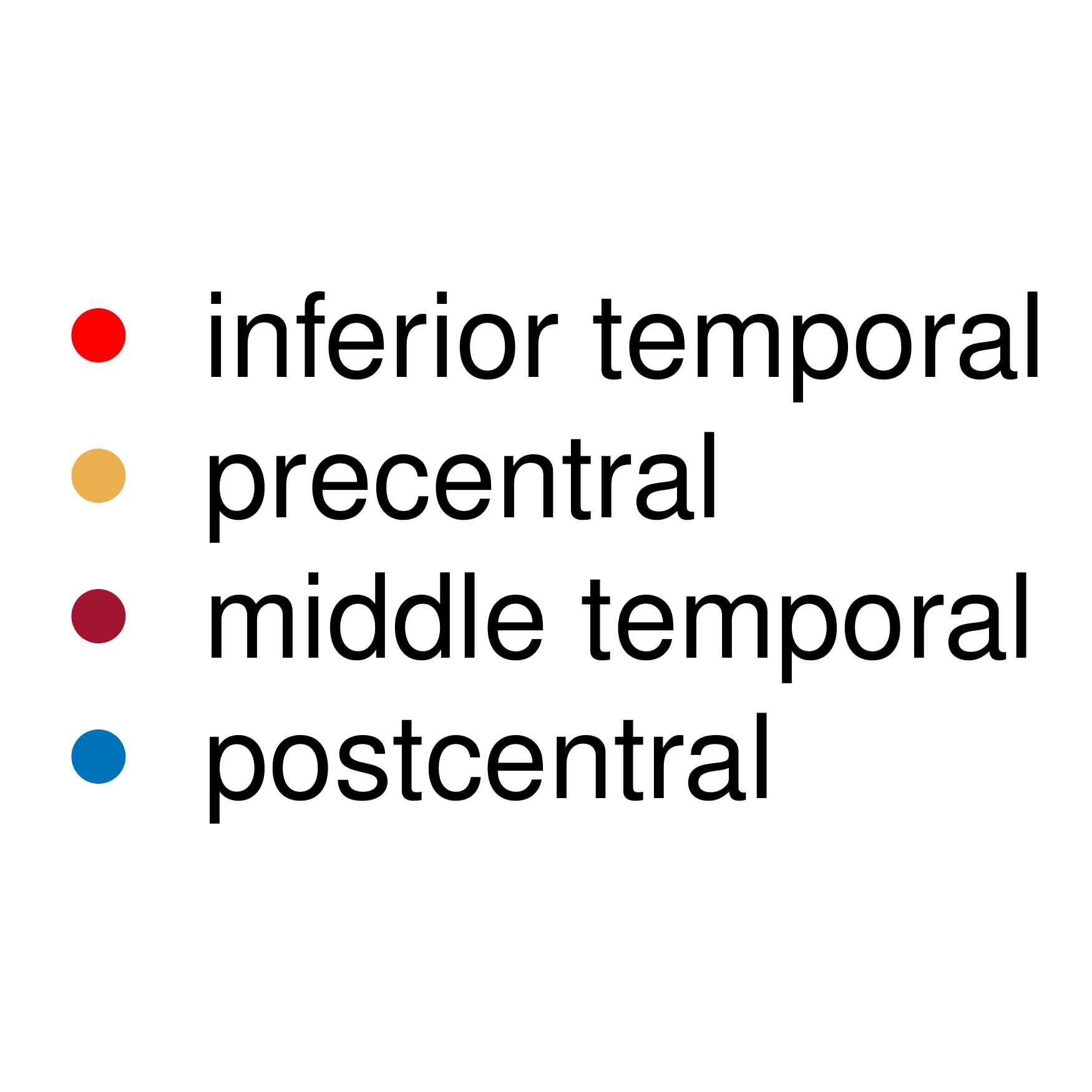}
\end{center}
\caption{\label{fig:real_brain_language}The identified brain pathway related to language. (a) DTI: inferior temporal gyrus (Left) -- precentral gyrus (Left), (b) fMRI: middle temporal gyrus (Left) -- postcentral gyrus (Left).}
\end{figure}

The other pathway, through the structural connectivity between left inferior temporal gyrus (\texttt{IT\_L})  and left precentral gyrus (\texttt{preCentral\_L}), then through the functional connectivity between left middle temporal gyrus (\texttt{MT\_L}) and left postcentral gyrus. Figure~\ref{fig:real_brain_language} presents this pathway, which is suggestive of a language pathway. Inferior temporal gyrus, as part of the inferior longitudinal and the inferior occipito-frontal fasciculi, is crucial for semantic processing~\citep{mandonnet2007does} and responsible for word naming~\citep{race2013area}. Middle temporal gyrus is typically viewed as part the language networks~\citep{ficek2018effect}. 

\section{Discussion}
\label{sec:discuss}

In this article, we proposed a method of multimodal mediation analysis. It requires the ordering of modalities in the proposed mediation pathways, but it does not require the ordering of potential individual mediators within each modality. We defined three types of indirect pathway effects, and employed a lasso-type regularization for estimation. We studied both the asymptotic and empirical behavior of our method. 

The proposed method neatly fits within the context of brain imaging data for combining diffusion weighted MRI with functional MRI data. Abstracting the setting, the framework interrogates the idea that, across a sample of subjects, an exposure or treatment impacts neural wiring, and the consequent changes in wiring impacts brain functional activity, which in turn impacts behavior. As our model is agnostic to domain, one might also consider applying the same approach where an exposure is postulated to impact epigenetic measurements, then subsequently genomic measurements, such as RNA expression, and changes in behavior or clinical outcome. 

In our data analysis, by integrating structural and functional imaging, we have postulated mechanistic pathways, including ones related to working memory and language, that mediate sex related differences in language behavior. However, care must be taken, as any mechanistic interpretation would be highly dependent on a variety of modeling assumptions, for instance, the path analysis ordering and the linearity assumption. 

Our work also points to a number of potential extensions. In our analysis, the model vectorized the connectivity measures and did not exploit the symmetric and positive definite matrix structure. Recent developments in covariance regression \citep{sun2017store, zhao2018covariate} are potentially useful. Also, we did not consider cases where the exposure and / or the outcome are themselves high-dimensional. We plan to pursue these lines of work as our future research.





\bibliographystyle{biom} 
\bibliography{Bibliography}



\appendix
\renewcommand{\thefigure}{\Alph{figure}}
\renewcommand{\thetable}{\Alph{table}}
\renewcommand{\theequation}{A\arabic{equation}}

\section{Proofs}

\subsection{Proof of Lemma~\ref{lemma:convex}}

\begin{proof}
By Theorem 1 in \citet{zhao2016pathway}, if and only if when $\nu\geq 1/2$, $v(a,b)=|ab|+\nu(a^{2}+b^{2})$ is a convex function. This proves the convexity of the function.
    
The following proves the second part of the lemma.
\[
|\beta_{j}\theta_{j}|+\nu_{1}\left(\beta_{j}^{2}+\theta_{j}^{2}\right) = \frac{1}{2}\left( |\beta_{j}|+|\theta_{j}| \right)^{2} + \left( \nu_{1}-\frac{1}{2} \right)\left(\beta_{j}^{2}+\theta_{j}^{2}\right)
\]
\[
\sum_{j=1}^{p_{1}}\left\{ \frac{1}{2}\left(|\beta_{j}|+|\theta_{j}|\right)^{2} + \left(\nu_{1}-\frac{1}{2}\right)\left(\beta_{j}^{2}+\theta_{j}^{2}\right) \right\} \leq r_{1} \quad \Rightarrow \quad \sum_{j=1}^{p_{1}}\beta_{j}^{2}\leq\frac{r_{1}}{\nu_{1}}
\]
Analogously,
\begin{eqnarray*}
\sum_{k=1}^{p_{2}}\pi_{k}^{2} & \leq & \frac{r_{2}}{\nu_{2}}, \\
 \left(\sum_{j=1}^{p_{1}}\sum_{k=1}^{p_{2}}|\lambda_{jk}||\beta_{j}\pi_{k}|\right)^{2} & \leq & \left(\sum_{j=1}^{p_{1}}\sum_{k=1}^{p_{2}}|\lambda_{jk}|^{2}\right)\left(\sum_{j=1}^{p_{1}}\sum_{k=2}^{p_{1}}|\beta_{j}\pi_{k}|^{2}\right) \\
        & \leq & \left(\sum_{j=1}^{p_{1}}\sum_{k=1}^{p_{2}}|\lambda_{jk}|\right)^{2}\left\{ \sum_{j=1}^{p_{1}}\beta_{j}^{2}\left(\sum_{k=1}^{p_{2}}\pi_{k}^{2}\right) \right\} \\
        & \leq & r_{3}^{2}\left\{ \sum_{j=1}^{p_{1}}\beta_{j}^{2}\left(\frac{r_{2}}{\nu_{2}}\right) \right\} \\
        & \leq & r_{3}^{2}\frac{r_{1}r_{2}}{\nu_{1}\nu_{2}}.
    \end{eqnarray*}
    We finish the proof by setting $r_{1}\leq t_{1}$, $r_{2}\leq t_{2}$, and $r_{3}\leq t_{3}\sqrt{\nu_{1}\nu_{2}/r_{1}r_{2}}$.
\end{proof}

\subsection{Proof of Theorem~\ref{thm:MSIPE_est}}

\begin{proof}
Before we prove Theorem~\ref{thm:MSIPE_est}, we briefly comment that, a key difference of our method compared to the method in \citet{zhao2016pathway} is the pathway effect through both $M_{1}$ and $M_{2}$ is decomposed as a product of three parameters. Lemma~\ref{lemma:convex} introduces a convex relaxation of the three-way product regularization. Therefore, in the following, based on the fact that the mediation effect is decomposed in three components, we prove the consistency of each component under such a convex regularization.

(i) We first consider the case with penalties $P_{1}$ and $P_{3}$. The estimator $\hat{\boldsymbol{\Theta}}$ is the solution to the optimization problem, 
\begin{eqnarray*}
\text{minimize} && \frac{1}{2}\ell(\boldsymbol{\Theta}), \\ 
\text{such that} && \sum_{j=1}^{p_{1}}\left\{|\beta_{j}\theta_{j}|+\nu_{1}(\beta_{j}^{2}+\theta_{j}^{2})\right\}\leq c_{11}, \\
& & \sum_{k=1}^{p_{2}}\left\{|\zeta_{k}\pi_{k}|+\nu_{2}(\zeta_{k}^{2}+\pi_{k}^{2})\right\}\leq c_{12}, \;\; 
\sum_{j=1}^{p_{1}}\sum_{k=1}^{p_{1}}|\lambda_{jk}|\leq c_{3}.
\end{eqnarray*}
Let 
\begin{eqnarray*}
\mathcal{C}_{1} &=& \left\{X\boldsymbol{\beta}\boldsymbol{\theta}:P_{11}(\boldsymbol{\beta},\boldsymbol{\theta})\leq r_{1}\right\}, \quad \text{where} \;\; P_{11}(\boldsymbol{\beta},\boldsymbol{\theta})=\sum_{j=1}^{p_{1}}\left\{|\beta_{j}\theta_{j}|+\nu_{1}(\beta_{j}^{2}+\theta_{j}^{2})\right\}, \\
\mathcal{C}_{2} &=& \left\{X\boldsymbol{\zeta}\boldsymbol{\pi}:P_{12}(\boldsymbol{\zeta},\boldsymbol{\pi})\leq r_{2}\right\}, \quad \text{where} \;\; P_{12}(\boldsymbol{\zeta},\boldsymbol{\pi})=\sum_{k=1}^{p_{2}}\left\{|\zeta_{k}\pi_{k}|+\nu_{2}(\zeta_{k}^{2}+\pi_{k}^{2})\right\}, \\
\mathcal{C}_{3} &=& \left\{X\boldsymbol{\beta}\boldsymbol{\Lambda}\boldsymbol{\pi}:P_{11}(\boldsymbol{\beta},\boldsymbol{\theta})\leq r_{1},P_{12}(\boldsymbol{\zeta},\boldsymbol{\pi})\leq r_{2},P_{3}(\boldsymbol{\Lambda})\leq r_{3}\right\}, \\
\mathcal{C} &=& \left\{V_{1}+V_{2}+V_{3}:V_{1}\in\mathcal{C}_{1},V_{2}\in\mathcal{C}_{2},V_{3}\in\mathcal{C}_{3}\right\}.
\end{eqnarray*}
Let $V_{1}=\bM_{1}\boldsymbol{\theta}=X\boldsymbol{\beta}^{*}\boldsymbol{\theta}^{*}+\boldsymbol{\epsilon}\boldsymbol{\theta}^{*}$, $V_{2}=X\boldsymbol{\zeta}^{*}\boldsymbol{\pi}^{*}+\boldsymbol{\vartheta}\boldsymbol{\pi}^{*}$, and $V_{3}=\bM_{1}\boldsymbol{\Lambda}^{*}\boldsymbol{\pi}^{*}+\xi=X\boldsymbol{\beta}^{*}\boldsymbol{\Lambda}^{*}\boldsymbol{\pi}^{*}+\boldsymbol{\epsilon}\boldsymbol{\Lambda}^{*}\boldsymbol{\pi}^{*}+\xi$. Then $V=V_{1}+V_{2}+V_{3}$. Note that $\hat{V}_{1}$, $\hat{V}_{2}$ and $\hat{V}_{3}$ are the projections of $V_{1}$, $V_{2}$, $V_{3}$ onto $\mathcal{C}_{1}$, $\mathcal{C}_{2}$ and $\mathcal{C}_{3}$, respectively. We have, 
\begin{eqnarray*}
\|\hat{V}-V^{*}\|_{2}^{2} &=& \|(\hat{V}_{1}-V_{1}^{*})+(\hat{V}_{2}-V_{2}^{*})+(\hat{V}_{3}-V_{3}^{*})\|_{2}^{2} \\
&\leq& \|\hat{V}_{1}-V_{1}^{*}\|_{2}^{2}+\|\hat{V}_{2}-V_{2}^{*}\|_{2}^{2}+\|\hat{V}_{3}-V_{3}^{*}\|_{2}^{2}.
\end{eqnarray*}
Next we bound the expectation of $\|\hat{V}_{l}-V_{l}^{*}\|_{2}^{2}$, $l=1,2,3$, respectively. 

For $\mathbb{E}\|\hat{V}_{1}-V_{1}^{*}\|_{2}^{2}$, we note that, for any $ x\in\mathcal{C}_{1}$, $\langle V_{1}-\hat{V}_{1}, x-\hat{V}_{1} \rangle \leq 0$. Setting $x=V_{1}^{*}$, then
\begin{eqnarray*}
        \|\hat{V}_{1}-V_{1}^{*}\|_{2}^{2} &=& \langle \hat{V}_{1}-V_{1}^{*},\hat{V}_{1}-V_{1}^{*} \rangle \\
        &=& \langle \hat{V}_{1}-V_{1},\hat{V}_{1}-V_{1}^{*} \rangle + \langle V_{1}-V_{1}^{*},\hat{V}_{1}-V_{1}^{*} \rangle \\
        &\leq& \langle V_{1}-V_{1}^{*},\hat{V}_{1}-V_{1}^{*} \rangle.
\end{eqnarray*}
By Definition~\ref{def:prediction}, we have, 
\begin{eqnarray*}
        \hat{V}_{1}-V_{1}^{*} &=& X\hat{\boldsymbol{\beta}}\hat{\boldsymbol{\theta}}-X\boldsymbol{\beta}^{*}\boldsymbol{\theta}^{*}=X(\hat{\boldsymbol{\beta}}\hat{\boldsymbol{\theta}}-\boldsymbol{\beta}^{*}\boldsymbol{\theta}^{*}), \\
        V_{1}-V_{1}^{*} &=& (X\boldsymbol{\beta}^{*}\boldsymbol{\theta}^{*}+\boldsymbol{\epsilon}\boldsymbol{\theta}^{*})-X\boldsymbol{\beta}^{*}\boldsymbol{\theta}^{*}=\boldsymbol{\epsilon}\boldsymbol{\theta}^{*}.
\end{eqnarray*}
Therefore, 
\begin{eqnarray*}
\|\hat{V}_{1}-V_{1}^{*}\|_{2}^{2} \leq \sum_{i=1}^{n}\left(\boldsymbol{\epsilon}_{i}\boldsymbol{\theta}^{*}\right)\left\{X_{i}(\hat{\boldsymbol{\beta}}\hat{\boldsymbol{\theta}}-\boldsymbol{\beta}^{*}\boldsymbol{\theta}^{*})\right\}=(\hat{\boldsymbol{\beta}}\hat{\boldsymbol{\theta}}-\boldsymbol{\beta}^{*}\boldsymbol{\theta}^{*})\sum_{j=1}^{p_{1}}\theta_{j}^{*}\left(\sum_{i=1}^{n}\epsilon_{ij}X_{i}\right).
\end{eqnarray*}
Let $Q_{1j}=\sum_{i=1}^{n}\epsilon_{ij}X_{i}$. By \citet[Lemma 3]{chatterjee2013assumptionless}, we have, 
\begin{eqnarray*}
Q_{1j}\sim\mathcal{N}\left(0,\sigma_{1j}^{2}\sum_{i=1}^{n}X_{i}^{2}\right), \quad \text{and} \quad \mathbb{E}\left(\max_{1\leq j\leq p_{1}}|Q_{1j}|\right)\leq \sqrt{\sum_{i=1}^{n}X_{i}^{2}}\left(\max_{j}\sigma_{1j}\right)\sqrt{2\log(2p_{1})},
\end{eqnarray*}
where $\mathrm{diag}(\boldsymbol{\Sigma}_{1})=\{\sigma_{11}^{2},\dots,\sigma_{1p_{1}}^{2}\}$. Under the conditions $|X_{i}|\leq c_{0}$ and $\max_{j}\sigma_{1j}\leq c_{4}$, we have
\[
        \mathbb{E}\left(\max_{1\leq j\leq p_{1}}|Q_{1j}|\right)\leq c_{0}c_{4} \sqrt{2n\log(2p_{1})}.
\]
In addition, $P_{11}(\hat{\boldsymbol{\beta}},\hat{\boldsymbol{\theta}})\leq c_{11}$ and $P_{11}(\boldsymbol{\beta}^{*},\boldsymbol{\theta}^{*})\leq c_{11}$, then
\[
(\hat{\boldsymbol{\beta}}\hat{\boldsymbol{\theta}}-\boldsymbol{\beta}^{*}\boldsymbol{\theta}^{*})\leq P_{11}(\hat{\boldsymbol{\beta}},\hat{\boldsymbol{\theta}})+P_{11}(\boldsymbol{\beta}^{*},\boldsymbol{\theta}^{*}) \leq 2c_{11}.
\]
As $|\theta_{j}^{*}|\leq c_{01}$, we have, 
\[
        \mathbb{E}\|\hat{V}_{1}-V_{1}^{*}\|_{2}^{2}\leq 2c_{11}s_{2}c_{01}c_{0}c_{4}\sqrt{2n\log(2p_{1})}.
\]

For $\mathbb{E}\|\hat{V}_{2}-V_{2}^{*}\|_{2}^{2}$, we can bound it in an analogous way. That is, denote $\mathrm{diag}(\boldsymbol{\Sigma}_{2})=\{\sigma_{21}^{2},\dots,\sigma_{2p_{2}}^{2}\}$, $\max_{k}\sigma_{2k}\leq c_{4}$, and $|\pi_{k}^{*}|\leq c_{02}$, we have, 
\begin{eqnarray*} 
\|\hat{V}_{2}-V_{2}^{*}\|_{2}^{2} & \leq & \langle V_{2}-V_{2}^{*},\hat{V}_{2}-V_{2}^{*} \rangle = (\hat{\boldsymbol{\zeta}}\hat{\boldsymbol{\pi}}-\boldsymbol{\zeta}^{*}\boldsymbol{\pi}^{*})\sum_{k=1}^{p_{2}}\pi_{k}^{*}\left(\sum_{i=1}^{n}\vartheta_{ik}X_{i}\right), \\
\mathbb{E}\|\hat{V}_{2}-V_{2}^{*}\|_{2}^{2} & \leq & 2c_{12}s_{4}c_{02}c_{0}c_{4}\sqrt{2n\log(2p_{2})}.
\end{eqnarray*}

For $\mathbb{E}\|\hat{V}_{3}-V_{3}^{*}\|_{2}^{2}$, we have, 
\begin{eqnarray*}
\|\hat{V}_{3}-V_{3}^{*}\|_{2}^{2} &\leq& \langle V_{3}-V_{3}^{*},\hat{V}_{3}-V_{3}^{*} \rangle \\
        &=& (\hat{\boldsymbol{\beta}}\hat{\boldsymbol{\Lambda}}\hat{\boldsymbol{\pi}}-\boldsymbol{\beta}^{*}\boldsymbol{\Lambda}^{*}\boldsymbol{\pi}^{*})\sum_{i=1}^{n}(\xi_{i}X_{i})+(\hat{\boldsymbol{\beta}}\hat{\boldsymbol{\Lambda}}\hat{\boldsymbol{\pi}}-\boldsymbol{\beta}^{*}\boldsymbol{\Lambda}^{*}\boldsymbol{\pi}^{*})\sum_{j=1}^{p_{1}}\varsigma_{j}^{*}\left(\sum_{i=1}^{n}\xi_{ij}X_{i}\right).
\end{eqnarray*}
Let $\tilde{Q}_{j}=\sum_{i=1}^{n}\xi_{ij}X_{i}$. By \citet[Lemma 3]{chatterjee2013assumptionless}, we have, 
\[
\tilde{Q}_{j}\sim\mathcal{N}\left(0,\sigma^{2}\sum_{i=1}^{n}X_{i}^{2}\right), \quad \text{and} \quad \mathbb{E}\left(\max_{1\leq j\leq p_{1}}|\tilde{Q}_{j}|\right)\leq \sqrt{\sum_{i=1}^{n}X_{i}^{2}}\sigma\sqrt{2\log(2p_{1})}\leq c_{0}c_{4}\sqrt{2n\log(2p_{1})}.
\]
Under the condition that $|\varsigma_{j}^{*}|\leq c_{03}$, we have 
\[
\mathbb{E}\|\hat{V}_{3}-V_{3}^{*}\|_{2}^{2}\leq 2c_{3}\sqrt{c_{11}c_{12}/\nu_{1}\nu_{2}}(1+s_{5}c_{03})c_{0}c_{4}\sqrt{2n\log(2p_{2})},
\]
where $t_{3}=r_{3}\sqrt{r_{1}r_{2}/\nu_{1}\nu_{2}}$.
    
Putting the above three bounds together, we have, 
\begin{eqnarray*}
&& \mathbb{E}\left(\mathrm{MSPE}\right)
= \mathbb{E}\left\{\frac{1}{n}\sum_{i=1}^{n}(\hat{V}-V^{*})^2\right\} \\
        &\leq& 2c_{0}c_{4}\left\{c_{11}s_{2}c_{01}\sqrt{\frac{2\log(2p_{1})}{n}}+c_{12}s_{4}c_{02}\sqrt{\frac{2\log(2p_{2})}{n}}+c_{3}\sqrt{c_{11}c_{12}/\nu_{1}\nu_{2}}(1+s_{5}c_{03})\sqrt{\frac{2\log(2p_{1})}{n}}\right\}.
\end{eqnarray*}
 
Next, to establish the convergence of the pathway effect estimators in Theorem~\ref{thm:MSIPE_est}, we study the expectation of $(\hat{V}_{l}-V_{l}^{*})^{2}$, $l = 1, 2, 3$, respectively, 

For $\mathbb{E}(\hat{V}_{1}-V_{1}^{*})^{2}$, we have, 
\begin{eqnarray*}
\mathbb{E}\left(\hat{V}_{1}-V_{1}^{*}\right)^{2} & = & \mathbb{E}\left\{X(\hat{\boldsymbol{\beta}}\hat{\boldsymbol{\theta}}-\boldsymbol{\beta}^{*}\boldsymbol{\theta}^{*})\right\}^{2}=\mathbb{E}X^{2}\left(\hat{\boldsymbol{\beta}}\hat{\boldsymbol{\theta}}-\boldsymbol{\beta}^{*}\boldsymbol{\theta}^{*}\right)^{2}, \\
\frac{1}{n}\|\hat{V}_{1}-V_{1}^{*}\|_{2}^{2} & = & \frac{1}{n}\sum_{i=1}^{n}\left\{X_{i}(\hat{\boldsymbol{\beta}}\hat{\boldsymbol{\theta}}-\boldsymbol{\beta}^{*}\boldsymbol{\theta}^{*})\right\}^{2}=\left(\frac{1}{n}\sum_{i=1}^{n}X_{i}^{2}\right)\left(\hat{\boldsymbol{\beta}}\hat{\boldsymbol{\theta}}-\boldsymbol{\beta}^{*}\boldsymbol{\theta}^{*}\right)^{2}.
\end{eqnarray*}
Therefore, 
\begin{eqnarray*}
\mathbb{E}\left(\hat{V}_{1}-V_{1}^{*}\right)^{2}-\frac{1}{n}\|\hat{V}_{1}-V_{1}^{*}\|_{2}^{2}=\left(\mathbb{E}X^{2}-\frac{1}{n}\sum_{i=1}^{n}X_{i}^{2}\right)\left(\hat{\boldsymbol{\beta}}\hat{\boldsymbol{\theta}}-\boldsymbol{\beta}^{*}\boldsymbol{\theta}^{*}\right)^{2}.
\end{eqnarray*}
Let $Z_{i}=\mathbb{E}X^{2}-X_{i}^{2}$. Since $|X_{i}|\leq c_{0}$, $|Z_{i}|\leq 2c_{0}^{2}$, together with $\mathbb{E}X_{i}=0$, by  \citet[Lemma 5]{chatterjee2013assumptionless}, we have 
\[
        \mathbb{E}\left(e^{t\sum_{i=1}^{n}Z_{i}}\right)\leq e^{t^{2}n4c_{0}^{2}/2}.
\]
By \citet[Lemma 4]{chatterjee2013assumptionless},
\[
\mathbb{E}\left(|\sum_{i=1}^{n}Z_{i}|\right)\leq 2c_{0}^{2}\sqrt{2n\log(2)}, \quad \text{and} \quad \mathbb{E}\left(|\frac{1}{n}\sum_{i=1}^{n}Z_{i}|\right)\leq 2c_{0}^{2}\sqrt{\frac{2\log(2)}{n}}.
\]
Under condition (C3-1), $(\hat{\boldsymbol{\beta}}\hat{\boldsymbol{\theta}}-\boldsymbol{\beta}^{*}\boldsymbol{\theta}^{*})\leq 2c_{11}$. Then
\[
\mathbb{E}\left(\hat{V}_{1}-V_{1}^{*}\right)^{2}-\frac{1}{n}\|\hat{V}_{1}-V_{1}^{*}\|_{2}^{2}\leq 8c_{11}^{2}c_{0}^{2}\sqrt{\frac{2\log(2)}{n}},
\]
which implies that
\[
\mathbb{E}\left(\hat{V}_{1}-V_{1}^{*}\right)^{2}\leq 8c_{11}^{2}c_{0}^{2}\sqrt{\frac{2\log(2)}{n}}+2c_{11}s_{2}c_{01}c_{0}c_{4}\sqrt{\frac{2\log(p_{1})}{n}}.
\]

For $\mathbb{E}(\hat{V}_{2}-V_{2}^{*})^{2}$, we have, analogously,
\[
\mathbb{E}\left(\hat{V}_{2}-V_{2}^{*}\right)^{2}\leq 8c_{12}^{2}c_{0}^{2}\sqrt{\frac{2\log(2)}{n}}+2c_{12}s_{4}c_{02}c_{0}c_{4}\sqrt{\frac{2\log(p_{1})}{n}},
\]

For $\mathbb{E}(\hat{V}_{3}-V_{3}^{*})^{2}$, similarly, 
\[
        \mathbb{E}\left(\hat{V}_{3}-V_{3}^{*}\right)^{2}\leq 8c_{3}^{2}(c_{11}c_{12}/\nu_{1}\nu_{2})c_{0}^{2}\sqrt{\frac{2\log(2)}{n}}+2c_{3}\sqrt{c_{11}c_{12}/\nu_{1}\nu_{2}}(1+s_{5}c_{03})c_{0}c_{4}\sqrt{\frac{2\log(p_{1})}{n}}.
 \]

Together, the convergence of the pathway effect estimators in Theorem~\ref{thm:MSIPE_est} (i) follows. 

\medskip

(ii) We next consider the case with penalties $P_{1}$, $P_{2}$ and $P_{3}$. The proof is similar to the case with penalties $P_{1}$ and $P_{3}$.

Specifically, let 
\begin{eqnarray*}
\tilde{\mathcal{C}}_{1} & = & \left\{X\boldsymbol{\beta\theta}:P_{11}(\boldsymbol{\beta},\boldsymbol{\theta})\leq c_{11},P_{21}(\boldsymbol{\beta},\boldsymbol{\theta})\leq c_{21}\right\}, \quad \text{where} \;\; P_{21}(\boldsymbol{\beta},\boldsymbol{\theta})=\sum_{j=1}^{p_{1}}\left(|\beta_{j}|+|\theta_{j}|\right), \\
\tilde{\mathcal{C}}_{2} & = & \left\{X\boldsymbol{\zeta\pi}:P_{12}(\boldsymbol{\zeta},\boldsymbol{\pi})\leq c_{12},P_{22}(\boldsymbol{\zeta},\boldsymbol{\pi})\leq c_{22}\right\}, \quad \text{where} \;\; P_{22}(\boldsymbol{\zeta},\boldsymbol{\pi})=\sum_{k=1}^{p_{2}}\left(|\zeta_{k}|+|\pi_{k}|\right), \\ 
\tilde{\mathcal{C}}_{3} & = & \left\{X\boldsymbol{\beta\Lambda\pi}:P_{11}(\boldsymbol{\beta},\boldsymbol{\theta})\leq c_{11},P_{21}(\boldsymbol{\beta},\boldsymbol{\theta})\leq c_{21},P_{12}(\boldsymbol{\zeta},\boldsymbol{\pi})\leq c_{12},P_{22}(\boldsymbol{\zeta},\boldsymbol{\pi})\leq c_{22},P_{3}(\boldsymbol{\Lambda})\leq c_{3}\right\}, \\
\tilde{\mathcal{C}} & = & \left\{V_{1}+V_{2}+V_{3}:V_{1}\in\tilde{\mathcal{C}}_{1},V_{2}\in\tilde{\mathcal{C}}_{2},V_{3}\in\tilde{\mathcal{C}}_{3}\right\}.
\end{eqnarray*} 
Following similar arguments as case (i), we have, 
\begin{eqnarray*}
        \mathbb{E}\|\hat{V}_{1}-V_{1}^{*}\|_{2}^{2} &\leq& 2c_{11}c_{21}c_{0}c_{4}\sqrt{2n\log(2p_{1})}, \\
        \mathbb{E}\|\hat{V}_{2}-V_{2}^{*}\|_{2}^{2} &\leq& 2c_{12}c_{22}c_{0}c_{4}\sqrt{2n\log(2p_{2})}.
\end{eqnarray*}
Moreover, based on the fact that, 
\begin{eqnarray*}
        \sum_{j=1}^{p_{1}}\varsigma_{j}^{*}=\sum_{j=1}^{p_{1}}\left(\sum_{k=1}^{p_{2}}\lambda_{jk}^{*}\pi_{k}^{*}\right)\leq\left(\sum_{j=1}^{p_{1}}\sum_{k=1}^{p_{2}}|\lambda_{jk}^{*}|\right)\left(\sum_{k=1}^{p_{2}}|\pi_{k}^{*}|\right)\leq c_{3}c_{22},
\end{eqnarray*}
we have
\[
        \mathbb{E}\|\hat{V}_{3}-V_{3}^{*}\|_{2}^{2}\leq 2c_{3}\sqrt{c_{11}c_{12}/\nu_{1}\nu_{2}}(1+c_{3}c_{22})c_{0}c_{4}\sqrt{2n\log(2p_{2})}.
\]
Putting the above three bounds together, we have,
\begin{eqnarray*}
        && \mathbb{E}\left(\mathrm{MSPE}\right) \\
        &\leq& 2c_{0}c_{4}\left\{c_{11}c_{21}\sqrt{\frac{2\log(2p_{1})}{n}}+c_{12}c_{22}\sqrt{\frac{2\log(2p_{2})}{n}}+c_{3}\sqrt{c_{11}c_{12}/\nu_{1}\nu_{2}}(1+c_{3}c_{22})\sqrt{\frac{2\log(2p_{1})}{n}}\right\}.
\end{eqnarray*}

Next, we have that, 
 \begin{eqnarray*}
        \mathbb{E}\left(\hat{V}_{1}-V_{1}^{*}\right)^{2} &\leq& 8c_{11}^{2}c_{0}^{2}\sqrt{\frac{2\log(2)}{n}}+2c_{11}c_{21}c_{0}c_{4}\sqrt{\frac{2\log(p_{1})}{n}}, \\
        \mathbb{E}\left(\hat{V}_{2}-V_{2}^{*}\right)^{2} &\leq& 8c_{12}^{2}c_{0}^{2}\sqrt{\frac{2\log(2)}{n}}+2c_{12}c_{22}c_{0}c_{4}\sqrt{\frac{2\log(p_{1})}{n}}, \\
\mathbb{E}\left(\hat{V}_{3}-V_{3}^{*}\right)^{2} & \leq & 8c_{3}^{2}(c_{11}c_{12}/\nu_{1}\nu_{2})c_{0}^{2}\sqrt{\frac{2\log(2)}{n}} \\  
& & +2c_{3}\sqrt{c_{11}c_{12}/\nu_{1}\nu_{2}}(1+c_{3}c_{22})c_{0}c_{4}\sqrt{\frac{2\log(p_{1})}{n}}.
\end{eqnarray*}
Then the convergence of the pathway effect estimators in Theorem~\ref{thm:MSIPE_est} (ii) follows. 
\end{proof}

\section{Optimization}
\label{appendix:sec:ADMM}

We derive here the explicit forms of the estimators in Algorithm~\ref{alg:ADMM}. Consider the augmented Lagrangian function, 
\begin{eqnarray*}
\mathcal{L}(\boldsymbol{\Upsilon},\boldsymbol{\Lambda},\delta,\tilde{\boldsymbol{\Upsilon}}) = \frac{1}{2}\ell(\boldsymbol{\Upsilon},\boldsymbol{\Lambda},\delta)+ P_{1}(\tilde{\boldsymbol{\Upsilon}})+ P_{2}(\tilde{\boldsymbol{\Upsilon}})+ P_{3}(\boldsymbol{\Lambda},\delta)+\sum_{r=1}^{4}\left( \langle h_{r}(\boldsymbol{\Upsilon},\tilde{\boldsymbol{\Upsilon}}),\boldsymbol{\tau}_{r}\rangle+\frac{\rho}{2}\| h_{r}(\boldsymbol{\Upsilon},\tilde{\boldsymbol{\Upsilon}})\|_{2}^{2}\right),
\end{eqnarray*}
where $\boldsymbol{\Upsilon}=(\boldsymbol{\beta},\boldsymbol{\theta},\boldsymbol{\zeta},\boldsymbol{\pi})$, 
$\tilde{\boldsymbol{\Upsilon}}=(\tilde{\boldsymbol{\beta}},\tilde{\boldsymbol{\theta}},\tilde{\boldsymbol{\zeta}},\tilde{\boldsymbol{\pi}})$, 
$h_{1}(\boldsymbol{\Upsilon},\tilde{\boldsymbol{\Upsilon}}) = \boldsymbol{\beta}-\tilde{\boldsymbol{\beta}}$, 
$h_{2}(\boldsymbol{\Upsilon},\tilde{\boldsymbol{\Upsilon}}) = \boldsymbol{\theta}-\tilde{\boldsymbol{\theta}}$, 
$h_{3}(\boldsymbol{\Upsilon},\tilde{\boldsymbol{\Upsilon}}) = \boldsymbol{\zeta}-\tilde{\boldsymbol{\zeta}}$,
$h_{4}(\boldsymbol{\Upsilon},\tilde{\boldsymbol{\Upsilon}}) = \boldsymbol{\pi}-\tilde{\boldsymbol{\pi}}$, 
$\boldsymbol{\tau}_{1},\boldsymbol{\tau}_{2}\in\mathbb{R}^{p_{1}}$, and $\boldsymbol{\tau}_{3},\boldsymbol{\tau}_{4}\in\mathbb{R}^{p_{2}}$.

For $\boldsymbol{\Upsilon}=(\boldsymbol{\beta},\boldsymbol{\theta},\boldsymbol{\zeta},\boldsymbol{\pi})$, letting $\bW_{1}=\boldsymbol{\mathrm{I}}_{p_{1}}$, $\bW_{2}=\boldsymbol{\mathrm{I}}_{p_{2}}$ and $w=1$, we have,  
\begin{eqnarray*}
    \frac{\partial\mathcal{L}}{\partial\boldsymbol{\beta}} &=& -\bX^\top(\bM_{1}-\bX\boldsymbol{\beta})+\boldsymbol{\tau}_{1}^\top+\rho(\boldsymbol{\beta}-\tilde{\boldsymbol{\beta}}), \\
    \frac{\partial\mathcal{L}}{\partial\boldsymbol{\theta}} &=& -\bM_{1}^\top(\bY-\bX\delta-\bM_{1}\boldsymbol{\theta}-\bM_{2}\boldsymbol\pi)+\boldsymbol{\tau}_{2}+\rho(\boldsymbol{\theta}-\tilde{\boldsymbol{\theta}}), \\
    \frac{\partial\mathcal{L}}{\partial\boldsymbol{\zeta}} &=& -\bX^\top(\bM_{2}-\bX\boldsymbol{\zeta}-\bM_{1}\boldsymbol{\Lambda})+\boldsymbol{\tau}_{3}^\top+\rho(\boldsymbol{\zeta}-\tilde{\boldsymbol{\zeta}}), \\
    \frac{\partial\mathcal{L}}{\partial\boldsymbol{\pi}} &=& -\bM_{2}^\top(\bY-\bX\delta-\bM_{1}\boldsymbol{\theta}-\bM_{2}\boldsymbol{\pi})+\boldsymbol{\tau}_{4}+\rho(\boldsymbol{\pi}-\tilde{\boldsymbol{\pi}}).
\end{eqnarray*}
Therefore, we have
\begin{eqnarray*}
    \boldsymbol{\beta} &=& (\bX^\top\bX+\rho)^{-1}(\bX^\top\bM_{1}-\boldsymbol{\tau}_{1}^\top+\rho\tilde{\boldsymbol{\beta}}) \\
    \boldsymbol{\theta} &=& (\bM_{1}^\top\bM_{1}+\rho\boldsymbol{\mathrm{I}})^{-1}\left\{\bM_{1}^\top(\bY-\bX\delta-\bM_{2}\boldsymbol{\pi})-\boldsymbol{\tau}_{2}+\rho\tilde{\boldsymbol{\theta}}\right\} \\
    \boldsymbol{\zeta} &=&(\bX^\top\bX+\rho)^{-1}\left\{\bX^\top(\bM_{2}-\bM_{1}\boldsymbol{\Lambda})-\boldsymbol{\tau}_{3}^\top+\rho\tilde{\boldsymbol{\zeta}}\right\} \\
    \boldsymbol{\pi} &=& (\bM_{2}^\top\bM_{2}+\rho\boldsymbol{\mathrm{I}})^{-1}\left\{\bM_{2}^\top(\bY-\bX\delta-\bM_{1}\boldsymbol{\theta})-\boldsymbol{\tau}_{4}+\rho\tilde{\boldsymbol{\pi}}\right\}
\end{eqnarray*}

For $\delta$, we have 
\begin{eqnarray*}
\frac{\partial\mathcal{L}}{\partial\delta}=-\bX^\top(\bY-\bX\delta-\bM_{1}\boldsymbol{\theta}-\bM_{2}\boldsymbol{\pi})+\kappa_{4}\sgn{\delta}.
\end{eqnarray*}
Therefore, we have 
\begin{eqnarray*}
\quad \delta=\frac{1}{\bX^\top\bX}\mathcal{S}_{\kappa_{4}}\left\{\bX^\top(\bY-\bM_{1}\boldsymbol{\theta}-\bM_{2}\boldsymbol{\pi})\right\},
\end{eqnarray*}
where $\mathcal{S}_{\kappa}(\mu)=\max\{|\mu|-\kappa,0\}\sgn{\mu}$ is the soft-thresholding function.

For $\boldsymbol{\Lambda}$, it is equivalent to $p_{2}$ standard lasso problems. That is, for $k=1,\dots,p_{2}$, we seek to 
\begin{eqnarray*}
\text{minimize} \quad \frac{1}{2}\|\bM_{2k}-\bX\zeta_{j}-\bM_{1}\boldsymbol{\Lambda}_{k}\|_{2}^{2}+\kappa_{3}\|\boldsymbol{\Lambda}_{k}\|_{1}.
\end{eqnarray*}
This is equivalent to a Lasso problem with $(\bM_{2k}-\bX\zeta_{j})$ as the ``outcome" and $\bM_{1}$ as the ``predictor".

For $\tilde{\boldsymbol{\beta}}$ and $\tilde{\boldsymbol{\theta}}$, we seek to minimize the function,
\begin{eqnarray*}
\kappa_{1}\sum_{j=1}^{p_{1}}\left\{|\tilde{\beta}_{j}\tilde{\theta}_{j}|+\nu_{1}(\tilde{\beta}_{j}^{2}+\tilde{\theta}_{j}^{2})\right\}+\mu_{1}\sum_{j=1}^{p_{1}}\left(|\tilde{\beta}_{j}|+|\tilde{\theta}_{j}|\right)+(\boldsymbol{\beta}-\tilde{\boldsymbol{\beta}})\boldsymbol{\tau}_{1}+\frac{\rho}{2}\|\boldsymbol{\beta}-\tilde{\boldsymbol{\beta}}\|_{2}^{2}+\boldsymbol{\tau}_{2}^\top(\boldsymbol{\theta}-\tilde{\boldsymbol{\theta}})+\frac{\rho}{2}\|\boldsymbol{\theta}-\tilde{\boldsymbol{\theta}}\|_{2}^{2},
\end{eqnarray*}
which can be minimized one element at a time. That is, for $j=1,\dots,p_{1}$, we minimize
\begin{eqnarray*}
\kappa_{1}\left\{|\tilde{\beta}_{j}\tilde{\theta}_{j}|+\nu_{1}(\tilde{\beta}_{j}^{2}+\tilde{\theta}_{j}^{2})\right\}+\mu_{1}\left(|\tilde{\beta}_{j}|+|\tilde{\theta}_{j}|\right)+\tau_{1k}(\beta_{j}-\tilde{\beta}_{j})+\frac{\rho}{2}(\beta_{j}-\tilde{\beta}_{j})^{2}+\tau_{2j}(\theta_{j}-\tilde{\theta}_{j})+\frac{\rho}{2}(\theta_{j}-\tilde{\theta}_{j})^{2},
\end{eqnarray*}
which is equivalent to minimizing 
\begin{eqnarray*}
\kappa_{1}|\tilde{\beta}_{j}\tilde{\theta}_{j}|+\mu_{1}|\tilde{\beta}_{j}|+\mu_{1}|\tilde{\theta}_{j}|+\frac{1}{2}(2\kappa_{1}\nu_{1}+\rho)\tilde{\beta}_{j}^{2}+\frac{1}{2}(2\kappa_{1}\nu_{1}+\rho)\tilde{\theta}_{j}^{2}-(\tau_{1k}+\rho\beta_{j})\tilde{\beta}_{j}-(\tau_{2j}+\rho\theta_{j})\tilde{\theta}_{j}.
\end{eqnarray*}
This can be solved by \citet[Lemma 3.2]{zhao2016pathway}. 

For $\tilde{\boldsymbol{\zeta}}$ and $\tilde{\boldsymbol{\pi}}$, we seek to minimize the function,
\begin{eqnarray*}
\kappa_{2}\sum_{k=1}^{p_{2}}\left\{|\tilde{\zeta}_{k}\tilde{\pi}_{k}|+\nu_{2}(\tilde{\zeta}_{k}^{2}+\tilde{\pi}_{k}^{2})\right\}+\mu_{2}\sum_{k=1}^{p_{2}}\left(|\tilde{\zeta}_{k}|+|\tilde{\pi}_{k}|\right)+(\boldsymbol{\zeta}-\tilde{\boldsymbol{\zeta}})\boldsymbol{\tau}_{3}+\frac{\rho}{2}\|\boldsymbol{\zeta}-\tilde{\boldsymbol{\zeta}}\|_{2}^{2}+\boldsymbol{\tau}_{4}^\top(\boldsymbol{\pi}-\tilde{\boldsymbol{\pi}})+\frac{\rho}{2}\|\boldsymbol{\pi}-\tilde{\boldsymbol{\pi}}\|_{2}^{2},
\end{eqnarray*}
which can again be minimized one element at a time. That is, for $k=1,\dots,p_{2}$, we minimize
\begin{eqnarray*}
\kappa_{2}\left\{|\tilde{\zeta}_{k}\tilde{\pi}_{k}|+\nu_{2}(\tilde{\zeta}_{k}^{2}+\tilde{\pi}_{k}^{2})\right\}+\mu_{2}\left(|\tilde{\zeta}_{k}|+|\tilde{\pi}_{k}|\right)+\tau_{3j}(\zeta_{k}-\tilde{\zeta}_{k})+\frac{\rho}{2}(\zeta_{k}-\tilde{\zeta}_{k})^{2}+\tau_{4k}(\pi_{k}-\tilde{\pi}_{k})+\frac{\rho}{2}(\pi_{k}-\tilde{\pi}_{k})^{2},
\end{eqnarray*}
which is equivalent to minimizing
\begin{eqnarray*}
\kappa_{2}|\tilde{\zeta}_{k}\tilde{\pi}_{k}|+\mu_{2}|\tilde{\zeta}_{k}|+\mu_{2}|\tilde{\pi}_{k}|+\frac{1}{2}(2\kappa_{2}\nu_{2}+\rho)\tilde{\zeta}_{k}^{2}+\frac{1}{2}(2\kappa_{2}\nu_{2}+\rho)\tilde{\pi}_{k}^{2}-(\tau_{3j}+\rho\zeta_{k})\tilde{\zeta}_{k}-(\tau_{4k}+\rho\pi_{k})\tilde{\pi}_{k}. 
\end{eqnarray*}
This can again be solved by \citet[Lemma 3.2]{zhao2016pathway}. 

Finally, for $\boldsymbol{\tau}_{1},\boldsymbol{\tau}_{2},\boldsymbol{\tau}_{3},\boldsymbol{\tau}_{4}$, we have, 
\begin{eqnarray*}
\boldsymbol{\tau}_{1}^{(s+1)} &=& \boldsymbol{\tau}_{1}^{(s)}+\rho(\boldsymbol{\beta}-\tilde{\boldsymbol{\beta}})^\top, \\
\boldsymbol{\tau}_{2}^{(s+1)} &=& \boldsymbol{\tau}_{2}^{(s)}+\rho(\boldsymbol{\theta}-\tilde{\boldsymbol{\theta}}), \\
\boldsymbol{\tau}_{3}^{(s+1)} &=& \boldsymbol{\tau}_{3}^{(s)}+\rho(\boldsymbol{\zeta}-\tilde{\boldsymbol{\zeta}})^\top, \\
\boldsymbol{\tau}_{4}^{(s+1)} &=& \boldsymbol{\tau}_{4}^{(s)}+\rho(\boldsymbol{\pi}-\tilde{\boldsymbol{\pi}}). 
\end{eqnarray*}

\section{Simulations}
\label{appendix:sec:sim}

\subsection{Simulation setting}
\begin{figure}[h!]
\begin{center}
\subfloat[Model~\eqref{eq:LSEM_full}.]{\includegraphics[width=0.45\textwidth]{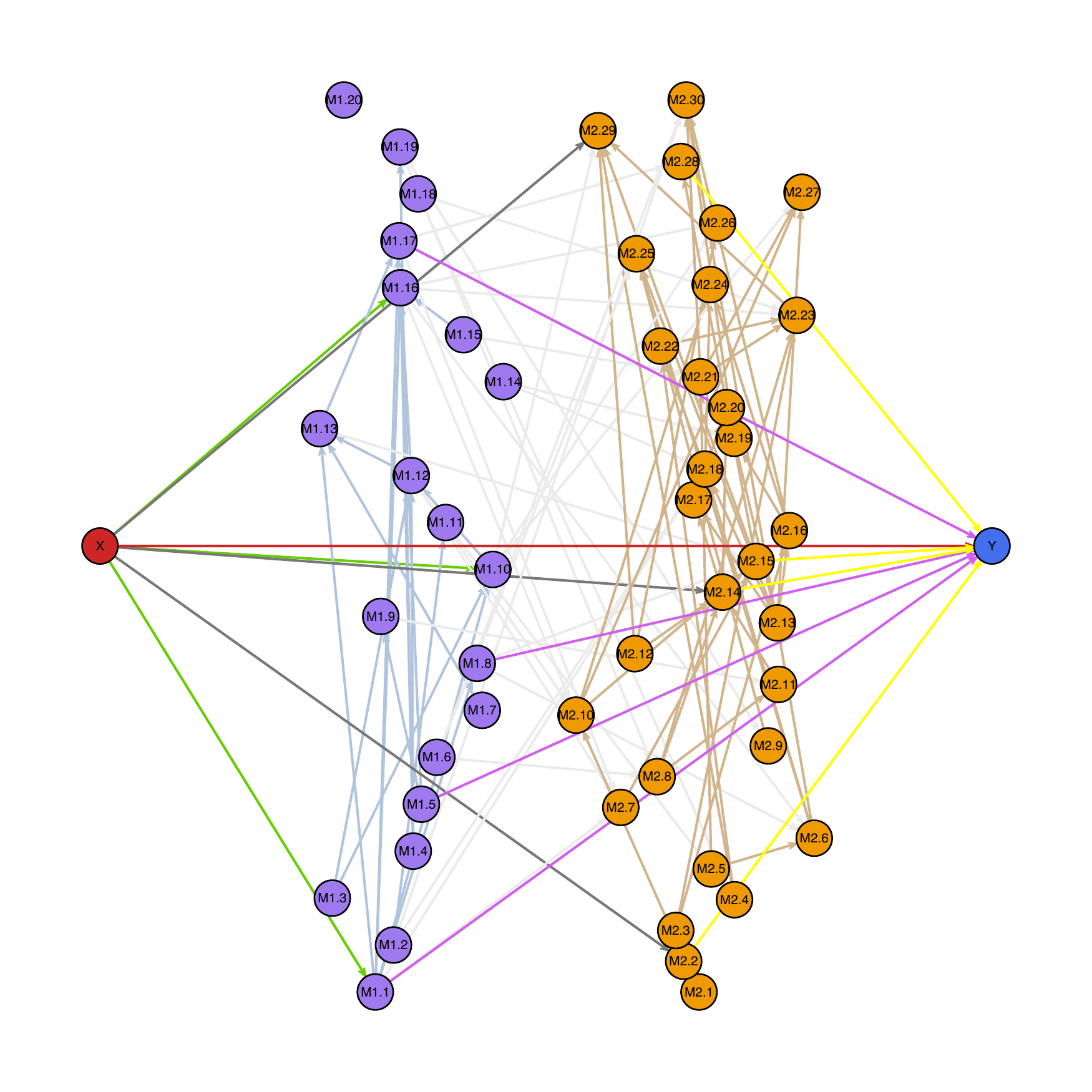}}
\enskip{}
\subfloat[Model~\eqref{eq:LSEM_reduced}]{\includegraphics[width=0.45\textwidth]{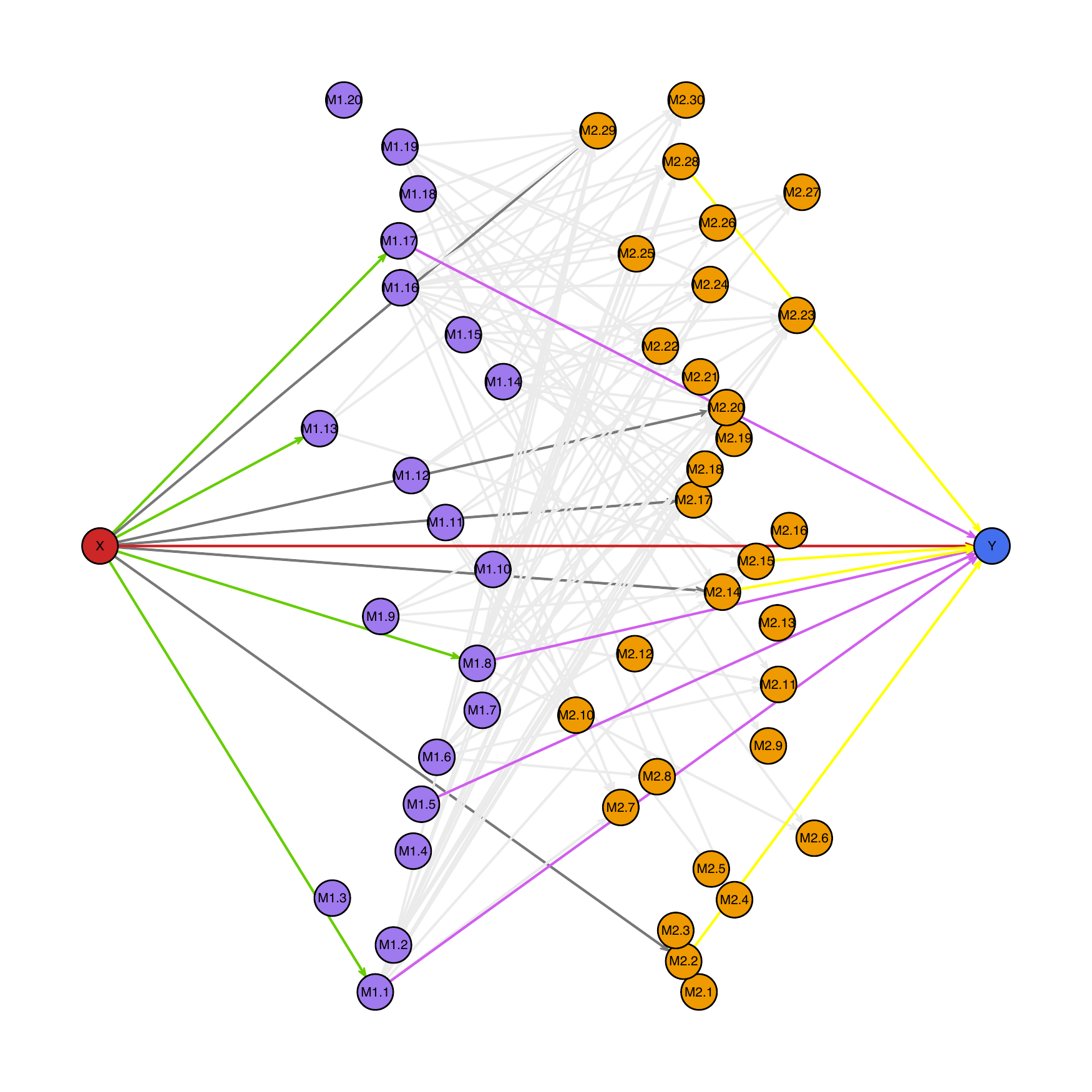}}
\end{center}
\caption{\label{appendix:fig:sim_setting}The generative scheme for the simulated data when $p_{1}=20$ and $p_{2}=30$.}
\end{figure}

Figure~\ref{appendix:fig:sim_setting} presents the generative scheme for the simulated data when $p_{1}=20$ and $p_{2}=30$.

\subsection{Additional results when $n=500$}

\begin{figure}[t!]
    \begin{center}
        \subfloat[ROC ($p_{1}=20, p_{2}=30$)]{\includegraphics[width=0.3\textwidth]{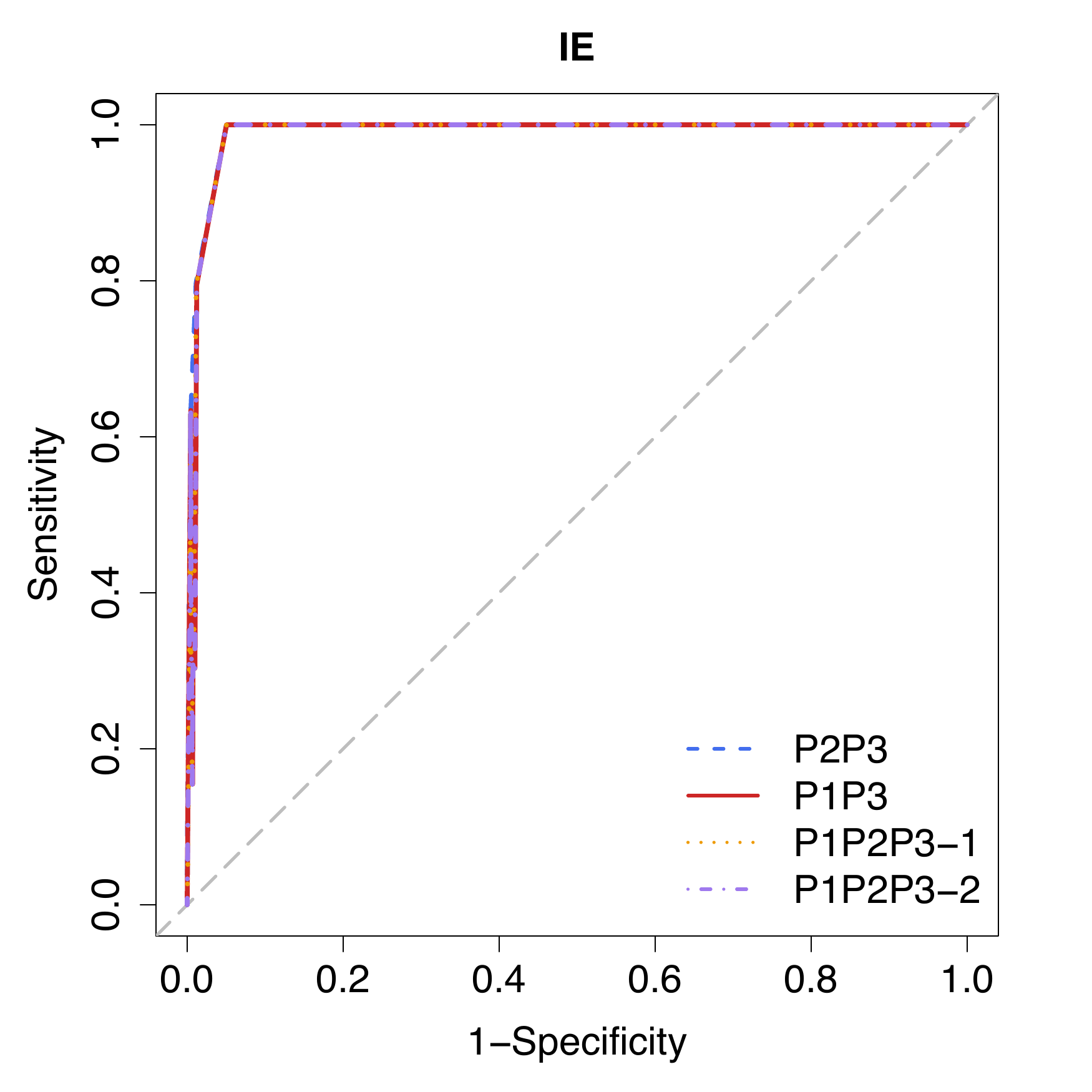}}
        \enskip{}
        \subfloat[MSE ($p_{1}=20, p_{2}=30$)]{\includegraphics[width=0.3\textwidth]{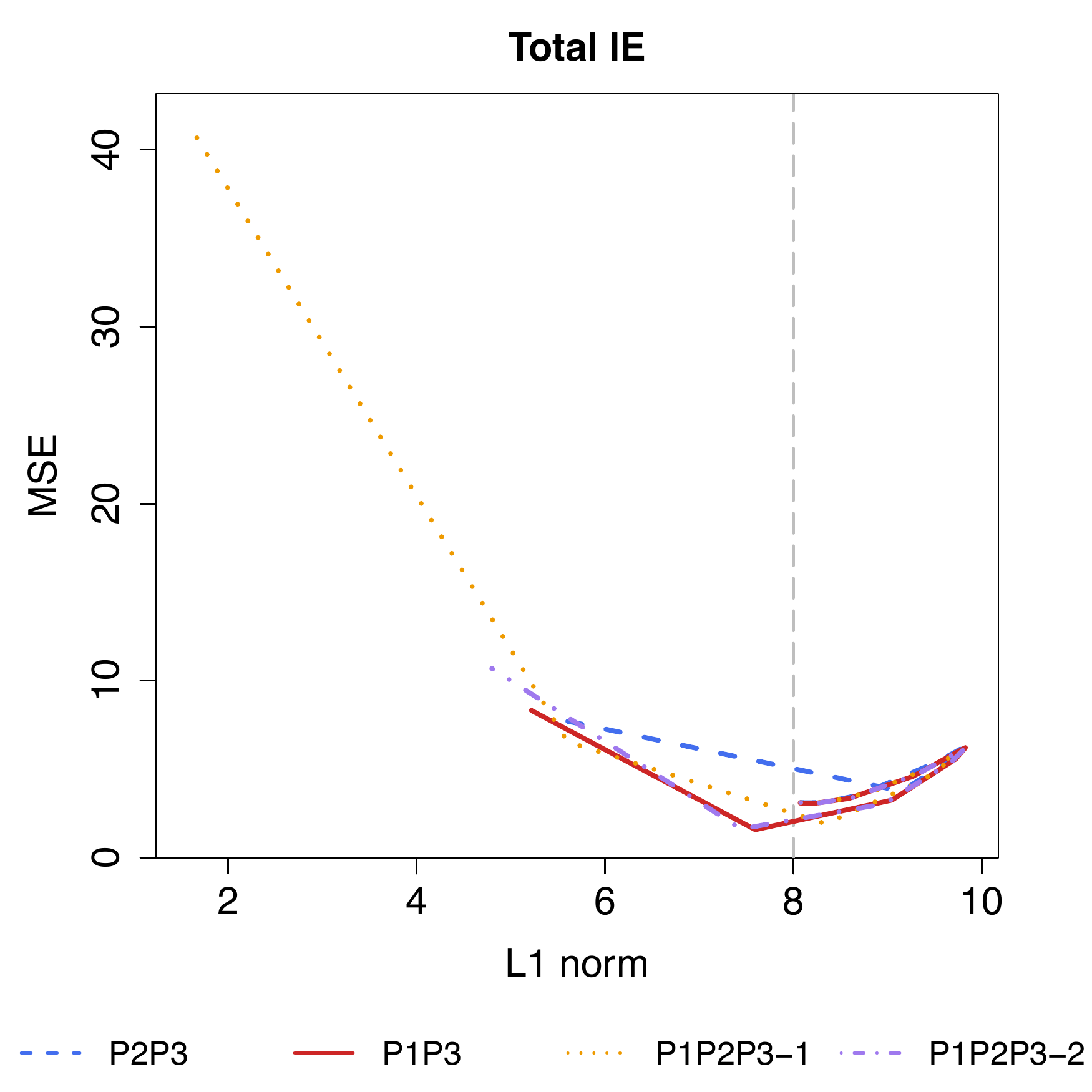}}
        \enskip{}
        \subfloat[Time ($p_{1}=20, p_{2}=30$)]{\includegraphics[width=0.3\textwidth]{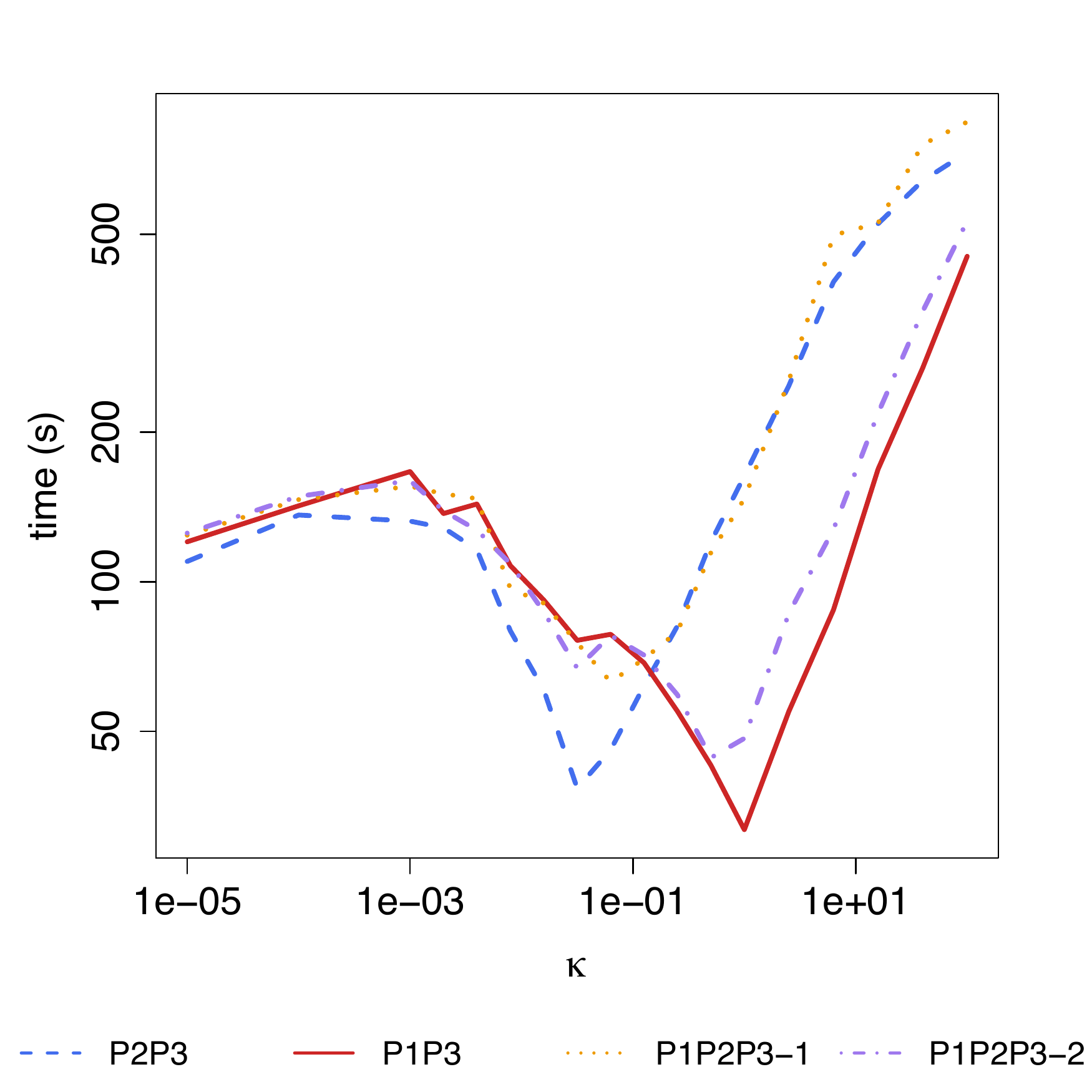}}
                
        \subfloat[ROC ($p_{1}=p_{2}=100$)]{\includegraphics[width=0.3\textwidth]{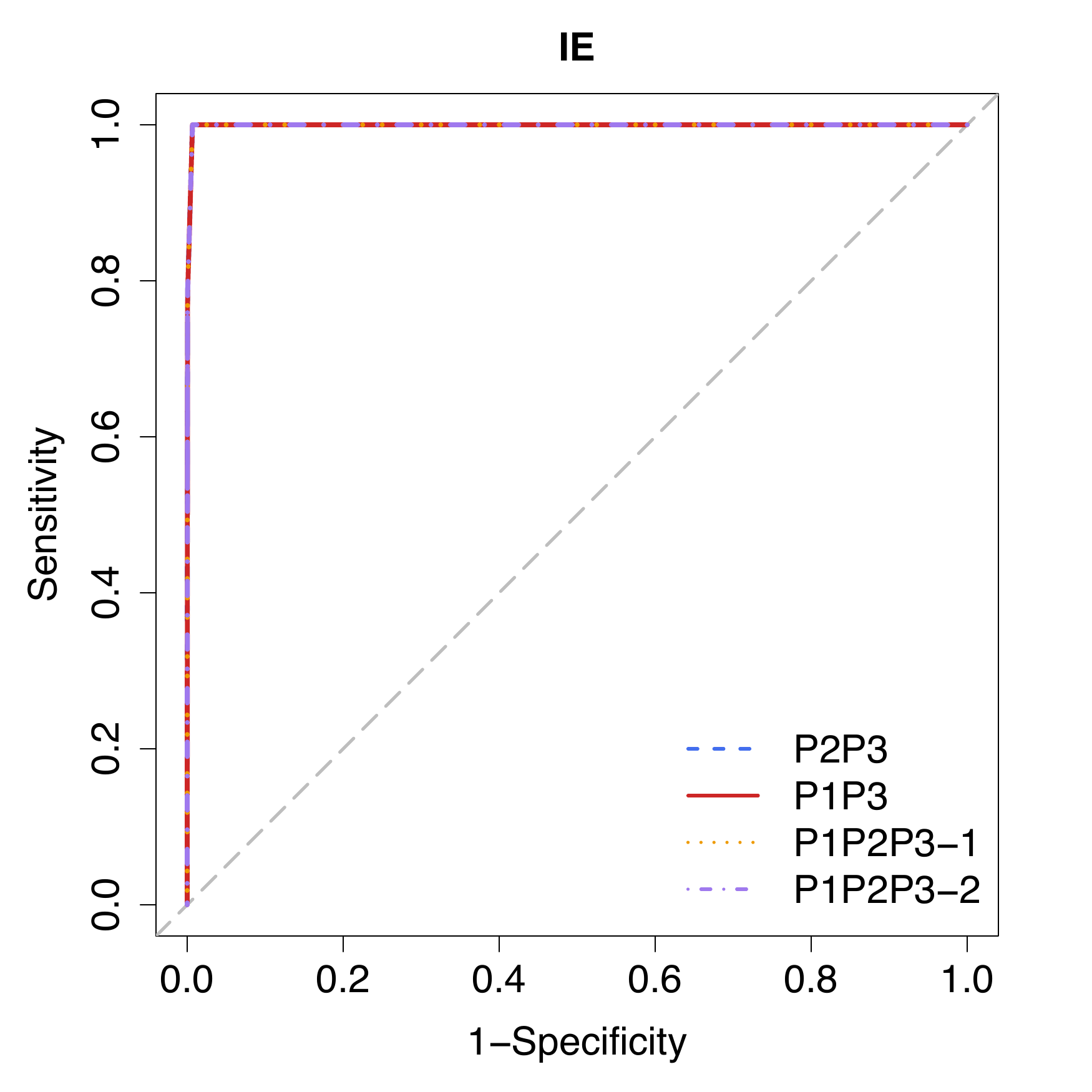}}
        \enskip{}
        \subfloat[MSE ($p_{1}=p_{2}=100$)]{\includegraphics[width=0.3\textwidth]{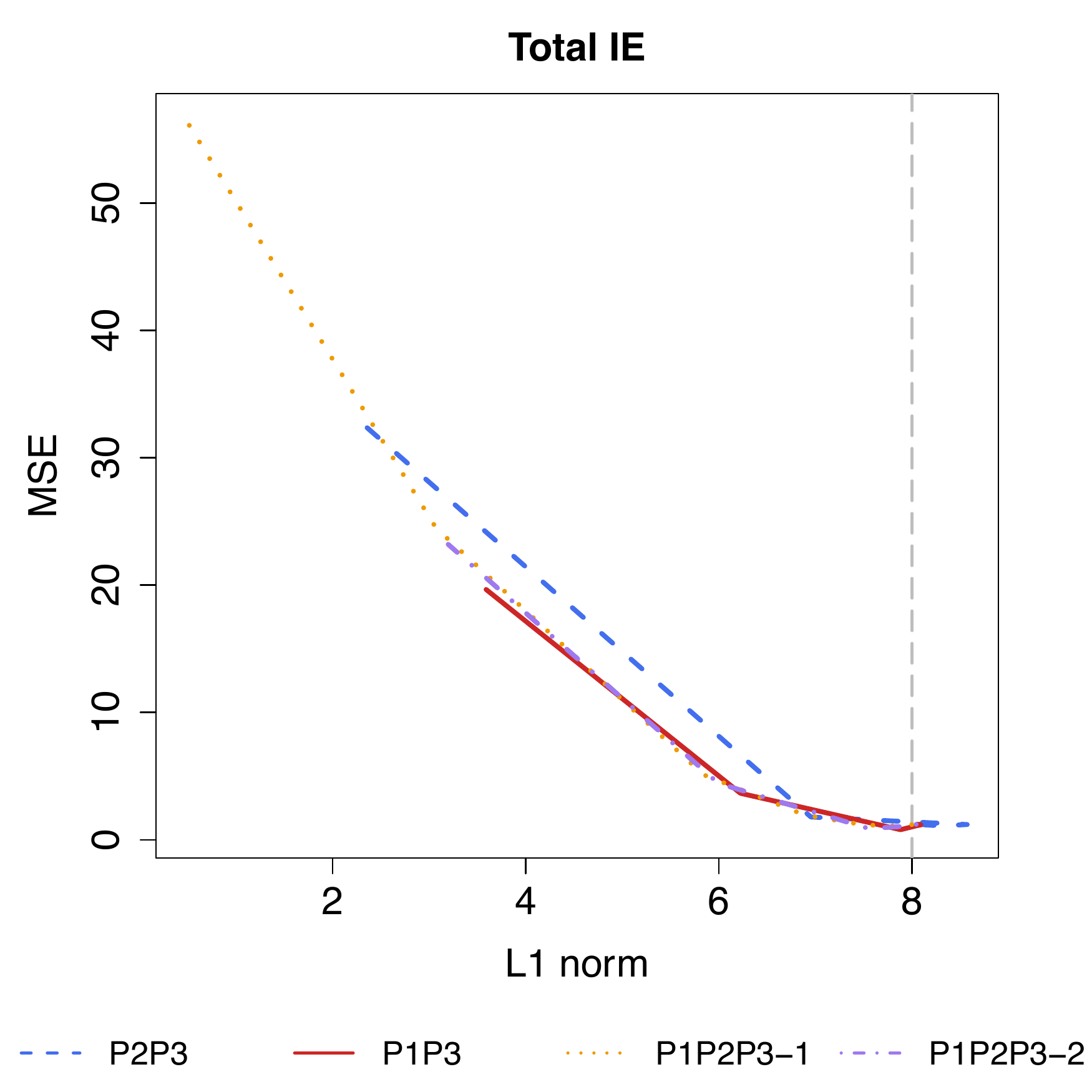}}
        \enskip{}
        \subfloat[Time ($p_{1}=p_{2}=100$)]{\includegraphics[width=0.3\textwidth]{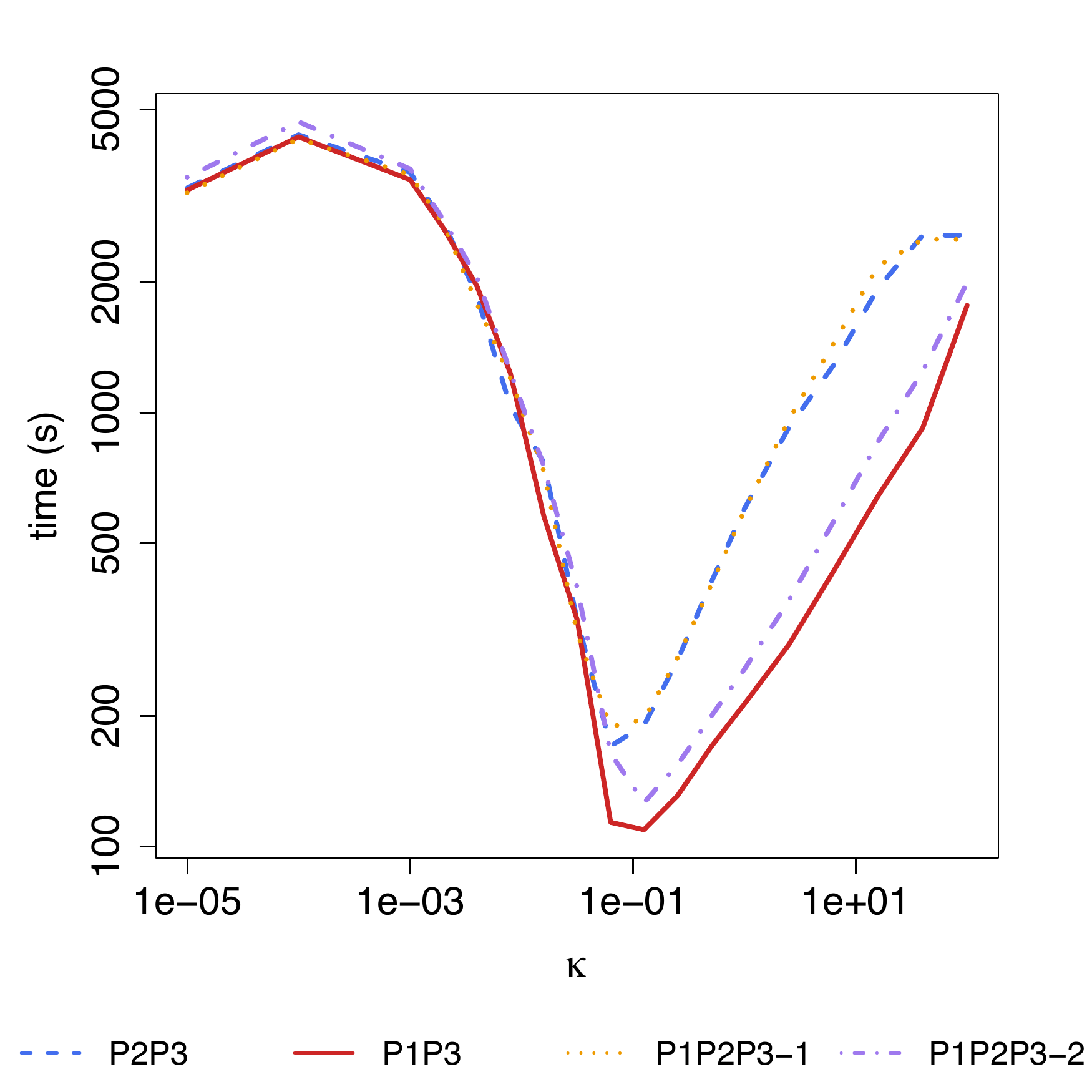}}
    \end{center}
    \caption{\label{appendix:fig:sim_n500}The estimate and the mean squared error of the total indirect effect, and the sensitivity and specificity of the indirect effect pathway selection, with the tuning parameter $\tilde\kappa$ selected by BIC. The sample size is $n=500$.}
\end{figure}

Figure~\ref{appendix:fig:sim_n500} and Table~\ref{appendix:table:sim_n500_BIC} present the average simulation results based on 200 data replications with $n=500$ and $p_{1}=p_{2}=100$. We see that, as the sample size increases, the performance of all methods improve and converge to the truth.

\begin{table}[b!]
\caption{\label{appendix:table:sim_n500_BIC}The estimate and the mean squared error of the total indirect effect, and the sensitivity and specificity of the indirect effect pathway selection, with the tuning parameter $\tilde\kappa$ selected by BIC. The sample size is $n=500$.}
    \begin{center}
        \begin{tabular}{l l R{2cm} R{2cm} R{2cm} R{2cm}}
            \hline
            & & \multicolumn{1}{c}{P2P3} & \multicolumn{1}{c}{P1P3} & \multicolumn{1}{c}{P1P2P3-1} & \multicolumn{1}{c}{P1P2P3-2} \\
            \hline
            & Truth & \multicolumn{4}{c}{8} \\
            & Estimate & 8.077 & 8.077 & 8.077 & 8.077 \\
            & MSE & 1.766 & 1.766 & 1.766 & 1.766 \\
            & Sensitivity & 0.568 & 0.568 & 0.568 & 0.568 \\
            \multirow{-5}{*}{$p_{1}=20,=p_{2}=30$} & Specificity & 0.996 & 0.996 & 0.996 & 0.996 \\
            \hline
            & Truth & \multicolumn{4}{c}{8} \\
            & Estimate & 8.027 & 8.027 & 8.027 & 8.027 \\
            & MSE & 1.223 & 1.223 & 1.223 & 1.223 \\
            & Sensitivity & 0.699 & 0.699 & 0.699 & 0.699 \\
            \multirow{-5}{*}{$p_{1}=p_{2}=100$} & Specificity & 0.999 & 0.999 & 0.999 & 0.999 \\
            \hline
        \end{tabular}
    \end{center}
\end{table}


\clearpage

\bibliographystyle{biom} 
\bibliography{Bibliography}

\end{document}